\documentclass[12pt,letterpaper]{JHEP3}
\usepackage{amsmath}
\usepackage{amssymb}
\usepackage{graphicx}
\usepackage{epsfig}
\usepackage{verbatim}
\advance\parskip 1.0pt plus 1.0pt minus 1.0pt   
\newcommand{\tr}{\operatorname{tr}}
\newcommand{\opname}[1]{\operatorname{#1}}

\newcommand{\sumintp}{\sum\negthickspace{\negthickspace{\negthickspace{\negthickspace{\negthickspace{\int_p^{\,\prime}}}}}}}
\newcommand{\Slash}[1]{{#1}\negthickspace{\negthickspace{\slash}}}

\def\Nc{N_c}
\def\CD{\half}
\def\Nfour{\mathcal{N}\,{=}\,4}
\def\half{{\textstyle \frac 12}}
\def\coeff#1#2{{\textstyle \frac {#1}{#2}}}

\preprint {}
\title
    {%
    Phase Diagram of \boldmath $\mathcal{N}\,{=}\,4$ Super-Yang-Mills Theory
    with $R$-Symmetry Chemical Potentials
    }%
\author
    {%
    Daisuke~Yamada\footnote{\tt dyamada@u.washington.edu}\;
    and Laurence~G. Yaffe\footnote{\tt yaffe@phys.washington.edu}
    \\Department of Physics, University of Washington, Seattle, WA 98195
    }%

\abstract
    {%
    The phase diagram of large $\Nc$, weakly-coupled $\Nfour$
    supersymmetric Yang-Mills theory on a three-sphere with non-zero
    chemical potentials is examined.
    In the zero coupling limit, a transition line in the $\mu$-$T$
    plane is found, separating a ``confined'' phase in which the Polyakov
    loop has vanishing expectation value from a ``deconfined'' phase in which
    this order parameter is non-zero.
    For non-zero but weak coupling, perturbative methods may be used
    to construct a dimensionally reduced effective theory valid
    for sufficiently high temperature.
    If the maximal chemical potential exceeds a critical value,
    then the free energy becomes unbounded below and
    no genuine equilibrium state exists.
    However, the deconfined plasma phase remains
    metastable, with a lifetime which grows exponentially with $\Nc$
    (not $\Nc^2$).
    This metastable phase persists with increasing chemical potential
    until a phase boundary,
    analogous to a spinodal decomposition line, is reached.
    Beyond this point, no long-lived locally stable quasi-equilibrium
    state exists.

    The resulting picture for the phase diagram of the weakly coupled
    theory is compared with results believed to hold in the strongly
    coupled limit of the theory, based on the AdS/CFT correspondence
    and the study of charged black hole thermodynamics.
    The confinement/deconfinement phase transition at weak coupling
    is in qualitative agreement with the Hawking-Page phase transition
    in the gravity dual of the strongly coupled theory.
    The black hole thermodynamic instability line
    may be the counterpart of
    the spinodal decomposition phase boundary
    found at weak coupling,
    but no black hole tunneling instability, analogous to
    the instability of the weakly coupled
    plasma phase is currently known.
    }%

\begin{document}

\section{Introduction}
The large $\Nc$ limit of a gauge theory \cite{'tHooft:1973jz}
is a type of classical limit \cite{Yaffe:1981vf}, and greatly simplifies
the structure of the theory.
Generalizing QCD from three to a large number $\Nc$ of colors
may appear to be a drastic modification, but the $\Nc=\infty$
theory exhibits many qualitative similarities to real QCD 
\cite{Veneziano:1976wm,Witten:1979kh}.
However, there are important differences as well.
For example, while real QCD (or any normal theory)
can only develop genuine phase transitions in the infinite volume limit,
$\Nc=\infty$ theories may have phase transitions even in finite volume,
because the $\Nc\to\infty$ limit acts as a thermodynamic limit.
One well-known example of this type is the Gross-Witten transition
in two-dimensional $U(\Nc)$ gauge theories \cite{Gross:1980he}.

More recently, large $\Nc$ gauge theories on a finite radius
three-sphere at non-zero temperature have been studied by Sundborg
\cite{Sundborg:1999ue},
and independently by Aharony \textit{et al.}
\cite{Aharony:2003sx,Aharony:2005bq}.
These theories were found to have phase transitions
even in the limit of zero gauge coupling.
Relevant order parameters include the Polyakov loop
and the dependence of free energy on $\Nc$.
In the high temperature phase, the Polyakov loop expectation value
is non-zero, the associated center symmetry is spontaneously broken,
and the free energy scales as $\Nc^2$ as $\Nc\to\infty$.
In the low temperature phase, the Polyakov loop has vanishing
expectation value, the center symmetry is unbroken,
and the free energy is order one with respect to $\Nc$
({\em i.e.}, the free energy has a finite $\Nc\to\infty$ limit).
In finite volume one cannot operationally define
confinement in terms of the free energy needed to separate
a static test quark and antiquark to infinity.
Nevertheless,
we will refer to these phases as ``deconfined'' and ``confining,''
respectively, because the corresponding phases of $SU(\Nc)$ Yang-Mills
theory in infinite volume show exactly the same features.
The reader should bear in mind that the justification for
this terminology comes only from the center
symmetry realization and the $\Nc$ dependence of the free energy.

In this paper we consider
$\Nfour$ supersymmetric Yang-Mills theory (SYM) with gauge group $SU(\Nc)$,
on a three-sphere
of radius $R$ and in the presence of non-zero chemical potentials
associated with the global $R$-symmetry.
Our goal is to understand the phase structure of the theory
as a function of both temperature and chemical potential.
We will focus on the weakly coupled limit of the theory,
but will compare results with those which are believed to
hold in the strongly coupled limit of the theory, based on the
study of solutions of the dual gravitational theory.

The outline of this paper is as follows.
In Section \ref{freetheory}, we consider the phase structure
of $\Nfour$ super-Yang-Mills theory on a sphere with
$R$-symmetry chemical potentials,
in the limit of zero gauge coupling.
Section \ref{hightemp} examines the theory with non-zero but weak coupling
in the high temperature regime ({\em i.e.}, inverse temperature
small compared to the radius of the spatial three-sphere).
In this regime, we construct a dimensionally reduced
effective field theory.
We compute terms in the resulting effective action up to fourth order
in the scalar fields, and evaluate the complete one-loop scalar field
effective potential in
the special case where the
fields take values in the flat directions of the tree-level potential.
Section~\ref{ESYM3thermodynamics} discusses the resulting
high temperature thermodynamics and, in particular,
the instability of the theory at 
sufficiently large chemical potential.
We summarize the results of our weak-coupling analysis 
in Section \ref{comparison} and compare with strong-coupling
results obtained from the dual gravitational theory.
Some possible extensions are discussed briefly in Section~\ref{outlook}.

Appendix \ref{MatrixModelDerivation} contains details
of the reduction of the zero coupling partition function
to the matrix model discussed in Section \ref{freetheory}.
Appendix \ref{Diagram} presents the one-loop diagrammatic calculations whose
results are summarized in Section \ref{hightemp}.
Appendix \ref{BGField} contains the details of the
evaluation of the one-loop scalar field effective potential,
using a background field method,
when the fields lie along flat directions.

The remainder of this introduction summarizes previous work on the
thermodynamics of $\Nfour$
super-Yang-Mills theory on a sphere.
We begin with strong-coupling results which emerge
from the AdS/CFT correspondence
(or gauge/string duality).
This correspondence,
which remains unproven but which has passed a great many consistency checks,
originated from the congruence of two apparently different theories:
Type IIB supergravity on AdS$_5\times S^5$ and large $\Nc$
$\mathcal{N}\,{=}\,4$ super-Yang-Mills theory
\cite{Maldacena:1997re,Gubser:1998bc}.
The field theory may be regarded as living 
on the $\mathbb{R}\times S^3$ boundary of AdS$_5$ space \cite{Witten:1998qj}.
The correspondence between bulk and boundary spaces extends 
to more general cases including ones where the boundary is $S^1\times S^3$.
The field theory that lives on this manifold is naturally viewed as a
finite temperature field theory,
with the temperature equaling the inverse of the $S^1$ circumference.

For an $S^1\times S^3$ boundary, there are two possible bulk geometries. 
One is (Euclidean signature) AdS$_5/\mathbb{Z}$
with the discrete group $\mathbb{Z}$ acting freely on the
AdS space, so the resulting manifold is smooth but topologically nontrivial.
This gives rise to two spin structures over the manifold.
The other bulk manifold with boundary $S^1\times S^3$
is the (Euclidean signature) Schwarzschild-AdS black hole.
This manifold is topologically trivial, and allows a single spin structure.
Thus, the sector of AdS$_5/\mathbb{Z}$ geometry which has the same spin
structure as the one in Schwarzschild-AdS may 
flop into Schwarzschild-AdS by a phase transition.
This is the Hawking-Page phase transition \cite{Hawking:1982dh}
and, as argued by Witten \cite{Witten:1998zw},
the signature of this transition in the dual thermal field theory
on $S^1 \times S^3$ should be a confinement/deconfinement phase transition.

Various systems that generalize this idea have also been studied.
These include,
on the gravity side, the thermodynamics of rotating black holes.
A rotation may be given either to the AdS$_5$ bulk or
to the internal $S^5$ sphere. The former is the Kerr-AdS black hole
and the latter, as seen from the bulk,
gives rise to the Reissner-Nordstr\"om-AdS 
(RN-AdS) black hole through the usual Kaluza-Klein mechanism whereby momenta
in compact extra dimensions appear as charges in the bulk.
It is this latter system which is of interest here.

\begin{FIGURE}[t]
{
\scalebox{.60}{\includegraphics{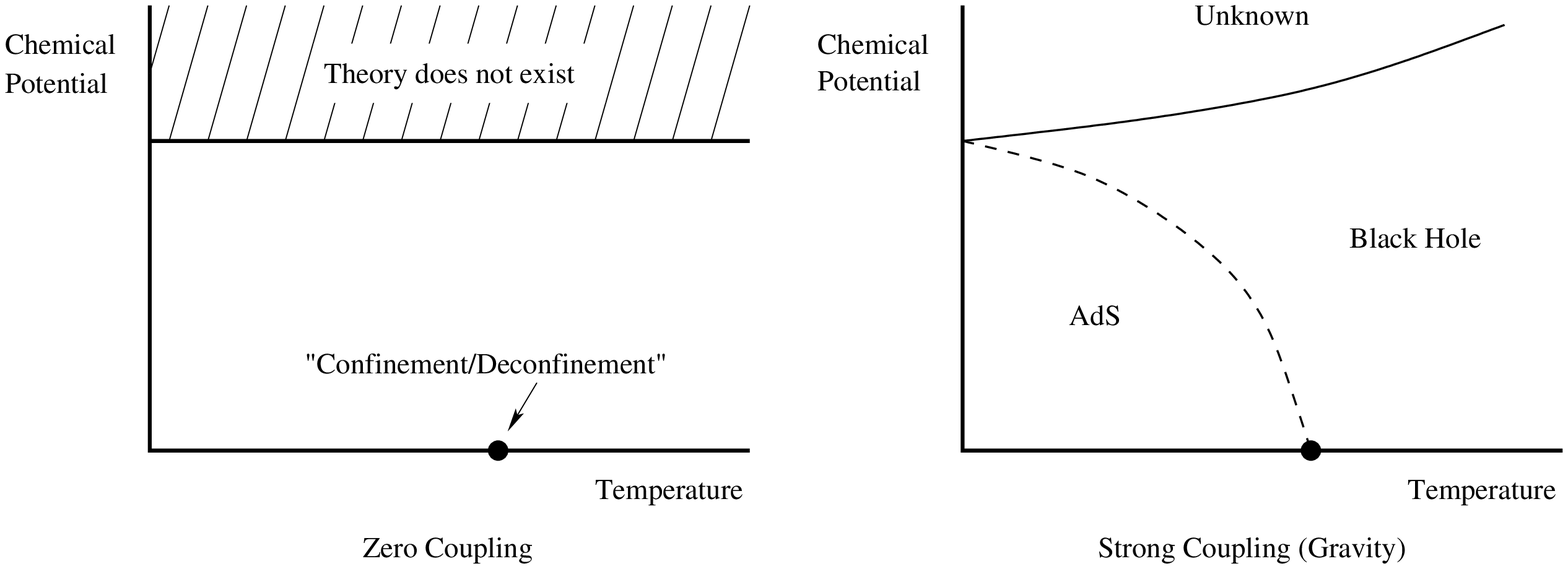}}
\caption
    {%
    Illustrations of previous results for the phase diagram
    of $\Nfour$ large $\Nc$ super-Yang-Mills theory as a function of
    temperature and chemical potential.
    The left hand figure depicts free field results from the
    zero coupling limit of the theory while
    the right hand figure comes from the analysis of the Einstein-Maxwell
    gravitational system (with three equal charges) which is
    believed to be dual to the strong coupling limit of the gauge theory.
    }%
\label{IntroPic}
}
\end{FIGURE}

The action that governs the
bulk thermodynamics is the five dimensional Einstein-Maxwell action
with negative cosmological constant (and Euclidean signature).%
\footnote{
  When the $S^5$ rotates equally in three independent
  planes, the resulting thermodynamics is described by the Einstein-Maxwell
  action.
  For the more general case of unequal rotations,
  the action is that of five dimensional $\mathcal{N}=8$ gauged supergravity.
  This is a generalization of the Einstein-Maxwell
  action and contains scalar fields which couple to the Maxwell
  fields. Solutions to the equations of motion have been obtained
  by Behrndt \textit{et al.} \cite{Behrndt:1998jd}; the resulting
  black hole solutions are similar to the RN-AdS solution of the
  Einstein-Maxwell action.
}
This action has two saddle
points. One is  AdS space without a black hole,
the other is the RN-AdS black hole.
One may use either the Reissner-Nordstr\"om charge,
or the associated chemical potential
to parameterize these solutions.%
\footnote{
  Due to the analytic continuation from Minkowski to Euclidean signature,
  the time component of the Maxwell field is pure imaginary,
  so the parallel transporter of the gauge field around the
  $S^1$ boundary circle is not a pure phase, but rather a real number
  whose logarithm equals the chemical potential divided by the temperature.
}
For the grand canonical ensemble, the relevant parameters are
temperature and chemical potential.
The equilibrium state of the system corresponds to whichever
of the two saddle point configurations minimizes the free energy
(which corresponds to the gravity action minus an appropriate boundary term
\cite{Gibbons:1976ue,Balasubramanian:1999re}).
The resulting phase diagram has a transition line separating
these two phases, drawn as the dashed line
in the right hand diagram of Fig.~\ref{IntroPic}.
This is the generalization of the Hawking-Page phase transition
to non-zero charge;
the original Hawking-Page transition is indicated in the figure by the
dot on the temperature axis at zero chemical potential.
This transition between AdS and RN-AdS black hole solutions
is a first order phase transition \cite{Chamblin:1999tk}.

Cvetic and Gubser analyzed the thermodynamic stability of the
black holes \cite{Cvetic:1999ne} and
found an instability line in the phase
diagram (in addition to the Hawking-Page phase transition line).
Beyond the instability line, the black hole that extremizes 
the Euclidean action becomes a saddle point;
this corresponds to a thermodynamic instability.
The instability line is indicated by the solid line
in the right hand diagram of Fig.~\ref{IntroPic}.%
\footnote
  {
  There are three possible conserved charges corresponding
  to three independent rotational planes of $S^5$.
  Fig.~\ref{IntroPic} illustrates the case with three 
  equal charges, where the supergravity action reduces to
  Einstein-Maxwell.
  (Ref.~\cite{Cvetic:1999ne} uses
  a different parametrization in plotting the phase diagram.
  After recasting their equations for the instability line
  in terms of temperature and chemical potential,
  one finds a result of the form shown in Fig.~\ref{IntroPic}.)
  For other charge configurations, such as a single non-zero
  charge, we find a phase diagram with similar
  general structure, but in which the Hawking-Page transition
  line meets the instability line at a non-zero temperature
  (where the horizon radius shrinks to zero).
  For these more general cases,
  the behavior of the system at temperatures
  below that corresponding to zero horizon radius is not
  currently understood.
  We will discuss this further in Section \ref{comparison}.
\label{fn:BHinstability}
}

A natural question is the fate
of the theory beyond this instability line.
For the case of the RN-AdS black hole in four dimensions,
Gubser and Mitra \cite{Gubser:2000mm} found that the instability point
is almost exactly where supergravity develops tachyonic modes.
(There is 0.7\% numerical discrepancy which is believed to be a
numerical analysis artifact.)
It is generally presumed
that the theory enters a new phase beyond the instability line.
But in the absence of any explicit solution corresponding to such
a putative new phase, it is also possible that this line represents
a genuine boundary to the phase diagram, beyond which no equilibrium
state exists.
This would be analogous to the situation at zero coupling, discussed below.
In short, the fate of the theory beyond the instability line is not
currently known.

We now turn to previous work on the thermodynamics of
weakly coupled $SU(\Nc)$ $\Nfour$ 
super-Yang-Mills theory with non-zero
chemical potentials for $R$-charges, in the large $\Nc$ limit.
If one considers the theory on flat space and introduces
any non-zero chemical potential, then there is an immediate problem ---
no ground state (or grand canonical equilibrium state) exists.
A chemical potential acts like a negative mass squared
for the scalar fields.
Due to conformal symmetry, this theory (on flat space) has no mass
term. Because there are flat directions in the super-potential, any non-zero
chemical potential immediately destabilizes the theory.

Compactifying space by replacing flat $\mathbb R^3$ with a three-sphere
of radius $R$ avoids this problem.
The coupling of the scalar fields to the spacetime curvature acts
like a positive mass squared and allows one to introduce
non-zero chemical potentials, provided they are sufficiently small.
In the limit of zero gauge coupling, the maximum chemical potential
equals the curvature-induced scalar mass, namely $1/R$.
For larger chemical potentials, the theory is again unstable
and no equilibrium state exists.
Hence, as illustrated on the left in Fig.~\ref{IntroPic},
in the free field theory the line $\mu = 1/R$
is a boundary of the phase diagram (just like the $T=0$ line)
and is not a phase transition line.

This point of instability
has been referred to as ``Bose-Einstein condensation'' in the
AdS/CFT literature (\textit{e.g.}, 
Refs.~\cite{Gubser:1998jb,Hawking:1999dp})
but this is highly mis-leading terminology.
In an interacting system, Bose-Einstein condensation corresponds
to spontaneous symmetry breaking of a global $U(1)$ symmetry.
In a $\mu$-$T$ phase diagram, a Bose-Einstein condensation line
is a phase transition line separating distinct phases ---
it is not a boundary of the phase diagram.

In the free theory on $S^3$ without a chemical potential
one finds \cite{Sundborg:1999ue,Aharony:2003sx},
a ``confinement/deconfinement'' phase transition
as the temperature is increased,
as noted earlier.
This is indicated by the dot on the temperature axis at zero chemical
potential on
the left side of Fig.~\ref{IntroPic}.

Hawking and Reall \cite{Hawking:1999dp}, considering the free theory
on $S^3$ with a chemical potential, noted the phase boundary at
$\mu = 1/R$, but did not find any confinement/deconfinement transition
and concluded that the free field theory in this particular system
is qualitatively different from the conjectured strongly coupled
dual gravity system.
However, these authors ignored the constraint of Gauss' law
and the associated requirement that physical states be gauge invariant.
It is the restriction to gauge invariant states which is responsible
for the confinement/deconfinement transition at vanishing chemical
potential \cite{Sundborg:1999ue,Aharony:2003sx}.
Maintaining Gauss' law, even at zero coupling,
is necessary to obtain valid results
for the limiting weak coupling behavior of the interacting theory.
In Section~\ref{freetheory} we reanalyze the free theory with
non-zero chemical potentials and the Gauss law constraint,
and show that there is a confinement/deconfinement phase transition
line which connects the $\mu = 0$, $T \ne 0$ transition point with the
zero temperature limit of stability point at $\mu = 1/R$ and $T = 0$.

From Fig.~\ref{IntroPic}, one may also note
differences in the instability/phase-boundary lines.
In the strong-coupling gravity dual,
the chemical potential grows with temperature on the instability line
\cite{Cvetic:1999ne} and
at asymptotically high temperatures,
the instability line rises
linearly.
In the free field theory, the limiting chemical potential
defining the phase diagram boundary
is independent of temperature and, as we show below,
is unaffected by the imposition of Gauss' law.
In Section~\ref{hightemp}
we consider the theory at non-zero but small coupling,
and high temperature, and find that a thermal mass is generated
in addition to the spatial curvature induced mass.
The resulting effective scalar mass (squared) remains positive until
one reaches a ``spinodal decomposition'' line on which
the chemical potential
also rises linearly at asymptotically high temperatures.
This is discussed further in section \ref{ESYM3thermodynamics}.

As this work was nearing completion, we became aware of the recent work
of Basu and Wadia \cite{Basu:2005pj}, which discusses the canonical
emsenble of $\Nfour$ super-Yang-Mills theory with fixed $R$-charges,
instead of fixed chemical potentials.
These authors examined an approximation to the zero coupling limit
of the gauge invariant partition function with a projection onto
states of a given $R$-charge, 
and found multiple saddle points with differing (but always non-zero)
expectation values for the Polyakov loop.
They also constructed a simple model illustrating the effect of turning on
a weak coupling, and discussed similarities between their model's results
and properties of $R$-charged anti-de-Sitter black holes.
At first sight, the results of Ref.~\cite{Basu:2005pj} appear quite
different from the picture of the $\Nfour$ phase diagram which
emerges from our analysis.
However, as we discuss in Section~\ref{canonical},
proper consideration of the relation between canonical and grand canonical
ensembles near a first order phase transition shows that the qualitative
results of Ref.~\cite{Basu:2005pj} are consistent with our phase structure.

\section{Zero Coupling Limit}\label{freetheory}

Consider free $\Nfour$ $SU(\Nc)$ supersymmetric Yang-Mills theory on $S^3$.
This field theory has 
a global $SU(4)$ $R$-symmetry (denoted $SU(4)_R$) and one may introduce 
chemical potentials coupled to the corresponding conserved charges.
It should be emphasized
that the ``free'' gauge theory considered in this section
is the zero coupling limit of the theory while retaining Gauss' law.
This is appropriate since Gauss' law holds
and physical states are necessarily gauge invariant at all non-zero values
of the coupling.
In particular,
one must have global charge neutrality of all physical states even with 
an infinitesimally small coupling constant.

For free gauge theories on a sphere, there are two equivalent techniques
for computing the resulting partition function:
counting gauge invariant states directly,
or using a suitable functional integral representation.
The former method, adopted in
Refs.~\cite{Sundborg:1999ue,Polyakov:2001af,Aharony:2003sx},
takes advantage of the conformal mapping
relating operators in the theory on $\mathbb{R}^4$ to
states of the theory on the sphere.
One counts the number of gauge invariant operators of a
given dimension (which maps to the energy of the state on the sphere),
and then sums over all such operators to obtain the partition function
on the sphere.

We choose to employ the functional integral approach
(also discussed in Ref.~\cite{Aharony:2003sx})
which represents the projection onto gauge invariant states
via an integral over the time component of the gauge field.
We begin,
in subsection \ref{chempreview},
by reviewing the introduction of chemical potentials for a maximal
Abelian subgroup of a non-Abelian global symmetry group,
and then in subsection \ref{chemplagrangian}
write down the Lagrangian of the theory
with the chemical potentials included.
Using this Lagrangian, subsection
\ref{PFunc} presents the matrix model which describes the partition
function of the free theory on the sphere.
We sketch the derivation of this matrix model, but leave details to
Appendix \ref{MatrixModelDerivation}. 
The resulting phase diagram for the free theory is discussed
in subsection \ref{phases}, and shown to have a
phase transition line separating a ``confining'' low temperature
or small chemical potential phase from a ``deconfined''
high temperature/large chemical potential phase.
In the final subsection \ref{1storder} we show that this is
a first order phase transition in the zero coupling limit,
but note that it is possible for the transition to
become continuous at any non-zero coupling.

\subsection{Chemical Potentials for Non-Abelian Symmetries}\label{chempreview}

Consider a system with Hamiltonian $\hat H$ and internal symmetry group 
$G$, assumed to be a semi-simple compact Lie group.
A Cartan subalgebra of $G$ is also a maximal Abelian subalgebra,
with dimension equal to the rank of the group $G$.
Compactness implies that any group element may be written as the
exponential of some element in the Lie algebra.
Let $\hat{U}(g)$ be the unitary operator representing an element $g$ of the
group, and define
\begin{equation}
	Z(\beta,g)=\tr\, [\hat{U}(g)e^{-\beta \hat H}] \;,
\end{equation}
where $\beta$ is the inverse temperature.
By assumption, $\hat U(g)$ commutes with $\hat H$ for all group elements $g$.
Consequently $Z(\beta,g)$ is a class function since
(using trace cyclicity),
\begin{equation}
	Z(\beta,g)
	=\tr\, [\hat{U}(\eta)\,\hat{U}(g)\,\hat{U}(\eta)^{-1}e^{-\beta H}]
	=\tr\, [\hat{U}(\eta g\eta^{-1})\,e^{-\beta H}]
	=Z(\beta,\eta g\eta^{-1})
							\;,
\end{equation}
for arbitrary $\eta$ in $G$.
Any group element $g$ is equivalent
(under conjugation by group elements)
to some element $h$ of a maximal Abelian subgroup.
And any element of a maximal Abelian subgroup may be
expressed as an exponential of a sum of
generators of a Cartan subalgebra,
$h = e^{i \gamma_p \hat{Q}_p}$ where
$\{\hat{Q}_p \,|\, p{=}1,\cdots, \opname{rank}(G)\}$
are the generators and $\{\gamma_p\}$ are real numbers.
We are adopting the convention that the generators 
$\{ \hat Q_p \}$ are Hermitian.
Therefore, $Z(\beta,g)$ may be regarded as a function of
the $\opname{rank}(G)$ real variables $\{\gamma_p\}$, and
rewritten as
\begin{equation}
	Z(\beta,\gamma_p)=\tr\, e^{-\beta \hat H}e^{i\gamma_p \hat Q_p}\;.
\end{equation}
After an analytic continuation $\gamma_p\rightarrow-i\beta\mu_p$,
this is the grand canonical partition function
\begin{equation}\label{generalGZ}
	Z(\beta,\mu_p)\equiv \tr\, \exp[-\beta(\hat H-\mu_p \,\hat Q_p)] \;,
\end{equation}
with chemical potentials $\{\mu_p\}$ associated with a maximal set
of commuting conserved charges $\{ \hat Q_p \}$.
This illustrates why, given a non-Abelian symmetry group,
it is natural to introduce chemical potentials
corresponding to a Cartan subalgebra of the group.%
\footnote
    {
    See,
    for example,
    Ref.~\cite{Haber:1981ts} for a
    different and perhaps more direct discussion that leads to the same
    conclusion.
    }

\subsection{$\Nfour$ Super-Yang-Mills Lagrangian with
Chemical Potentials}\label{chemplagrangian}

Four dimensional $\Nfour$ super-Yang-Mills theory,
in flat space, may be obtained from dimensional 
reduction of ${\cal N}\,{=}\,1$ super-Yang-Mills 
in ten dimensions \cite{Gliozzi:1976qd,Brink:1976bc}.
We consider this theory on $S^1\times S^3$,
and include chemical potentials associated with a $U(1)^3$ maximal
Abelian subgroup of the $SU(4)$ global $R$-symmetry.
As mentioned in the Introduction, this theory has conformal 
scalar-curvature coupling terms that will appear as mass terms
for the scalar fields.

To write the Lagrangian, we first determine how the
charges $\hat Q_p$ associated with the
chemical potentials transform each field of the theory.
The field content includes six scalars which we will
regard as components of an antisymmetric matrix,
$\phi_{ij}$ ($i,j=1,\ldots,4$) with $\phi_{ij}=-\phi_{ji}$,
four (left-handed) Weyl fermions $\lambda_i$, 
and a vector field $A_\nu$,
all transforming in the adjoint representation of the
gauge group $SU(\Nc)$.

The scalar fields $\phi_{ij}$ are complex,
and transform under the antisymmetric representation $\mathbf{6}$
of $SU(4)_R$.
One may repackage these fields as a
$\bar{\mathbf{6}}$, defined by
\begin{equation}
	\phi^{ij}\equiv \half\, \epsilon^{ijkl}\, \phi_{kl}
							\;.
\end{equation}
Since the $\bar{\mathbf 6}$ representation is the complex conjugate
of the $\mathbf{6}$,
it is consistent to impose the reality condition
$\phi^{ij}=\phi_{ij}^*$. More explicitly, this is
\begin{equation}\label{ComplexScalarRelations}
	\phi_{12}=\phi_{34}^*
							\;,\quad
	\phi_{13}=\phi_{42}^*
							\;,\quad
	\phi_{14}=\phi_{23}^*
							\;,
\end{equation}
and the reality constraint leaves six (times $\Nc^2-1$) real degrees of freedom
in the scalar fields.
To simplify later expressions, we will relabel the three independent
complex scalar fields as
\begin {equation}
    \phi_1 \equiv \phi_{12} \,,\qquad
    \phi_2 \equiv \phi_{13} \,,\qquad
    \phi_3 \equiv \phi_{14} \,.
\end {equation}
The counting of degrees of freedom,
along with transformation properties under the global $SU(4)_R$ symmetry,
are summarized in Table~\ref{table:dof}.
Note that the total number of degrees of freedom for bosons and fermions are
equal, as usual in supersymmetric theories.

\begin {TABLE}[ht]
{
\centerline{
\begin{tabular}{cc c}
Field & D.O.F. & $SU(4)_R$ \\ \hline
$\phi_p$ & $6\,(\Nc^2{-}1)$       & $\mathbf{6}$\\ 
$\lambda_i$ & $8\,(\Nc^2{-}1)$       & $\mathbf{4}$\\
$A_\nu$     & $2\,(\Nc^2{-}1)$       & $\mathbf{1}$
\end{tabular}}
\caption {\footnotesize
Counting of real degrees of freedom, and $SU(4)_R$ symmetry
representations, for the fields of $\Nfour$ super-Yang-Mills theory.}
\label{table:dof}
}
\end{TABLE}

We choose to represent the generators of the Cartan subalgebra,
in the fundamental representation $\mathbf{4}$, as
\begin{subequations}
\begin{align}
	Q^\mathbf{4}_1&=\half\opname{diag}(1,1,-1,-1)\;,\\
	Q^\mathbf{4}_2&=\half\opname{diag}(1,-1,1,-1)\;,\\
	Q^\mathbf{4}_3&=\half\opname{diag}(1,-1,-1,1)\;.
\end{align}
\end{subequations}
To obtain the corresponding form in the antisymmetric representation
$\mathbf{6}$, one may consider the
transformations of a rank-2 tensor under $\mathbf{4}\otimes\mathbf{4}$.
Assembling the six scalars into a six-component vector via
\begin{equation}
    \phi^T \equiv
    (\phi_1,\phi_1^*,\phi_2,\phi_2^*,\phi_3,\phi_3^*)^T \,,
\end{equation}
the Cartan generator matrices in this representation appear as
\begin{subequations}
\begin{align}
	Q^\mathbf{6}_1&=\opname{diag}(1, -1,0, 0, 0,0)\;,\\
	Q^\mathbf{6}_2&=\opname{diag}(0, 0, 1, -1, 0, 0)\;,\\
	Q^\mathbf{6}_3&=\opname{diag}(0, 0, 0,0, 1, -1)\;.
\end{align}
\end{subequations}
With these choices, one sees that
\begin{equation}
    \sum_{p=1}^3\> \mu_p \, Q_p^\mathbf{6}
    =
    \opname{diag}(\mu_1,-\mu_1,\mu_2,-\mu_2,\mu_3,-\mu_3)\;.
\end{equation}
Hence, $\sum_{p=1}^3 \mu_p \, \hat Q_p$ assigns eigenvalue
$+\mu_p$ to $\phi_p$ and
assigns $-\mu_p$ to the complex conjugates $\phi_p^*$.
For the fermions,
$
    \sum_{p=1}^3\mu_p \, Q_p^\mathbf{4}
    =
    \opname{diag}(\tilde\mu_1,\tilde\mu_2,\tilde\mu_3,\tilde\mu_4)
    $
with
\begin{subequations}
\begin{align}
	&\tilde{\mu}_1\equiv\half(\mu_1+\mu_2+\mu_3),\phantom- \quad
	 \tilde{\mu}_2\equiv\half(\mu_1-\mu_2-\mu_3),\; \\
	&\tilde{\mu}_3\equiv\half(-\mu_1+\mu_2-\mu_3),\quad
	 \tilde{\mu}_4\equiv\half(-\mu_1-\mu_2+\mu_3),\;
\end{align}%
\label{mu_tilde}%
\end{subequations}
so the fermion $\lambda_j$ has an effective chemical potential
$\tilde{\mu}_j$.

To derive the correct form of the action to use
in a functional integral
when chemical potentials are present, 
one could start from the Lagrangian without chemical potentials,
find the conserved charges of the global symmetry,
re-express these in terms of conjugate momenta,
use this to derive a Hamiltonian path integral representation
of the grand canonical partition function (\ref{generalGZ}),
and then integrate out the conjugate momenta.
But that is completely unnecessary.
It is equivalent, but simpler, to imagine gauging
the $U(1)^3$ Abelian global $R$-symmetry.
The time component of such a fictitious gauge field couples
to the conserved $U(1)^3$ $R$-charge densities,
just like the chemical potentials.
But the time component of a gauge field in a Euclidean
functional integral corresponds to $i$ times the Minkowski 
gauge potential.
Consequently, in a Euclidean functional integral,
adding chemical potentials is exactly equivalent to turning
on a constant imaginary value for the time component of a
background gauge field associated with the $U(1)^3$ global $R$-symmetry.
One ends up with a standard Euclidean
functional integral
in which the Lagrangian contains modified covariant derivatives for
the time direction,
\begin{equation}
	D_\nu\rightarrow D_\nu-\mu_p \, Q_p \, \delta_{\nu0} \;.
\end{equation}

It will be convenient for later use to rewrite the Weyl fermions
as Majorana fermions.
Recall that a massless two-component Weyl fermion $\lambda$,
in four dimensions,
may be converted to a four-component Majorana fermion
$\psi \equiv \binom{\lambda_\alpha}{\bar\lambda^{\dot{\alpha}}}$.
The corresponding terms in the Lagrange density are related via
\begin{equation}\label{eq:weyl to majorana}
	\lambda^\alpha \, (\tau_\nu)_{\alpha\dot{\beta}}\,
	    (\stackrel{\leftrightarrow}{D}_\nu-\tilde{\mu}\,\delta_{\nu,0})\,
	\bar{\lambda}^{\dot{\beta}}
	=\half\, \bar{\psi}\,(\Slash{D}-\tilde\mu\, \gamma_0\gamma_5)\,\psi
								\;,
\end{equation}
where we have defined
$\bar\psi \equiv (\lambda^\alpha, \bar\lambda_{\dot{\alpha}})$ and
\begin{subequations}
\begin{align}\label{defgamma}
	\tau_\nu&\equiv  (\mathbf{1},\, i\vec{\sigma})
							\;,\qquad
	\bar{\tau}_\nu \equiv  (\mathbf{1},-i\vec{\sigma})
							\;,\\
	\gamma_\nu&\equiv \begin{pmatrix}
			0 & \tau_\nu	\\
			\bar{\tau}_\nu & 0
		  \end{pmatrix}				\,,\quad
	\gamma_5 \equiv \gamma_0\gamma_1\gamma_2\gamma_3
	=
	\begin{pmatrix}
	    \, \mathbf{1} & \phantom- 0 \, \\ \, 0 & -\mathbf{1} \,
	\end{pmatrix} .
\end{align}
\end{subequations}
The Majorana spinors satisfy the condition
\begin{equation}\label{MajoranaCondition}
	\psi=C\bar{\psi}
				\;,
\end{equation}
where
$
	C=\left(
		\genfrac{}{}{0pt}{}{\epsilon_{\alpha\beta}}{0}
		\genfrac{}{}{0pt}{}{0}{\epsilon^{\dot{\alpha}\dot{\beta}}}
	\right)
$
is the charge conjugation matrix with
$\epsilon_{12} = -\epsilon_{21}\equiv-1$.

The gauge field $A_\mu$ may be regarded as a traceless Hermitian matrix.
Equivalently, it may be expanded as $A_\mu = A_\mu^a \, T^a$, with
real coefficients $\{A_\mu^a\}$ and Hermitian color generators
$T^a$ which satisfy
\begin{subequations}
\begin{align}
	[T^a,T^b]&=if^{abc}\,T^c \;,
	\\
\noalign{\hbox{and}}
	\tr\, (T^aT^b)&=\CD\,\delta^{ab} \;.
\label{CDnormalization}
\end{align}
\end{subequations}
With this choice, the structure constants $f^{abc}$ are real and completely 
antisymmetric.

The scalar and fermion fields may similarly be expanded in the basis
of color generators,
$
    \phi_p = \phi_p^a \, T^a
$
and
$
    \psi_i = \psi_i^a \, T^a
$.
The coefficient $\psi_i^a$ is a four-component Grassmann-valued spinor.
The conjugate spinor $\bar\psi_i^a$ is not independent,
but is related to $\psi_i^a$ via
the Majorana condition (\ref{MajoranaCondition}).
The coefficients $\phi_p^a$ satisfy the reality
condition (\ref {ComplexScalarRelations}) (for each $a$).
It will be convenient to introduce explicitly independent real
scalar fields via
$\phi_p \equiv (X_p+iY_p)/\sqrt{2}$
and to assemble these into a multiplet,
\begin {equation}
	\Phi \equiv (X_1,Y_1,X_2,Y_2,X_3,Y_3) \;.
\end {equation}
We will use a capital Latin index to denote components of this vector,
so $\Phi_A = \Phi_A^a \, T^a$ with $A = 1,\cdots,6$.

We are finally ready to write down the $\mathcal{N}=4$ SYM Lagrange
density
\cite{Gliozzi:1976qd,Brink:1976bc} in Euclidean signature
with the addition of chemical potentials and spatial curvature
induced mass terms:
\begin{align}\label{fullLagrangian}
	\mathcal{L}=&\tr\,\Bigl\{\half(F_{\mu\nu})^2
		+ (D_{\nu}X_p-i\mu_p\,\delta_{\nu,0} Y_p)^2
		+ (D_{\nu}Y_p+i\mu_p\,\delta_{\nu,0} X_p)^2
		+ R^{-2} \, (\Phi_A)^2
	\nonumber\\
		&+ i \bar{\psi}_i(\Slash{D}-\tilde \mu_i \, \gamma_0\gamma_5)\psi_i
		+\half \, g^2(i[\Phi_A,\Phi_B])^2
		-g\, \bar{\psi}_i
		\left[ (\alpha^p_{ij}X_p +i\beta^q_{ij}\gamma_5Y_q), \,\psi_j\right]
		\Bigr\}
	\,,
\end{align}
where
$
	F_{\mu\nu}=\partial_{\mu}A_{\nu}-\partial_{\nu}A_{\mu}
			+ig[A_{\mu},A_{\nu}]
$
and
$
	D_{\nu}=\partial_{\nu}+ig[A_{\nu},\,\cdot\,]
$.
The indices run over
$A,B=1,\cdots\!,6$, $p,q=1,\cdots\!,3$, $i,j=1,\cdots\!,4$,
and $a=1,\cdots\!,\Nc^2-1$
(with implied sums over all indices).
The $4\times 4$ matrices $\alpha^p$ and $\beta^q$ satisfy the relations
\begin{equation}
	\{\alpha^p,\alpha^q\}=-2\,\delta^{pq}\,\mathbf{1}_{4\times 4} \;,\quad	
	\{\beta^p,\beta^q\}=-2\,\delta^{pq}\,\mathbf{1}_{4\times 4} \;,\quad
	[\alpha^p,\beta^q]=0 \;,
\end{equation}
and explicit forms can be given as
\begin{subequations}
\begin{align}
	\alpha^1&=
	\begin{pmatrix}
		0 & \sigma_1	\\
		-\sigma_1 & 0
	\end{pmatrix}	,\;
	\alpha^2=
	\begin{pmatrix}
		0 & -\sigma_3	\\
		\sigma_3 & 0
	\end{pmatrix}	,\;
	\alpha^3=
	\begin{pmatrix}
		i\sigma_2 & 0	\\
		 0 & i\sigma_2
	\end{pmatrix}	,\;
				\\
	\beta^1&=
	\begin{pmatrix}
		0 & i\sigma_2	\\
		i\sigma_2 & 0
	\end{pmatrix}	,\;
	\beta^2=
	\begin{pmatrix}
		0 & \sigma_0	\\
		-\sigma_0 & 0
	\end{pmatrix}	,\;
	\beta^3=
	\begin{pmatrix}
		-i\sigma_2 & 0	\\
		 0 & i\sigma_2
	\end{pmatrix}	\,.
\end{align}
\end{subequations}
The partition function for the grand canonical ensemble
has the functional integral representation
\begin {equation}
    Z =
    \int {\cal D}A_\mu \> {\cal D}\psi_i \> {\cal D}\Phi_A \;
    e^{-\textstyle \!\int d^4x \> {\cal L} }\,,
\label {eq:Zint}
\end {equation}
where the integral is over fields taking values on
$S^3$ (of radius $R$)${} \times S^1$ (of radius $\beta$),
with the gauge and scalar fields periodic and the Grassmann-valued
fermion fields anti-periodic on the Euclidean time circle.

\subsection{Gauge Invariant Partition Function}\label{PFunc}

We want to compute the partition
function (\ref {eq:Zint}) in the free-field limit of the theory,
while preserving the projection onto gauge invariant states
which is embodied in the functional integral over $A_0$.
We sketch the calculation here, and leave details
to Appendix \ref{MatrixModelDerivation}.

Taking a naive $g \to 0$ limit of the Lagrangian (\ref {fullLagrangian})
is the wrong procedure;
this ignores the special role of $A_0$ as a Lagrange multiplier
enforcing Gauss' law and, as detailed below,
would lead to an ill-defined Gaussian integral due to a zero-mode
in the covariance operator for $A_0$.
The physically appropriate limit is obtained by
sending $g\to 0$ after
separating and rescaling the constant mode of $A_0$,
\begin {equation}
    A_0(x) \longrightarrow \tilde A_0(x) + a/g \,,
\label {eq:separate}
\end {equation}
with $\tilde A_0(x)$ constrained to have a vanishing spacetime integral,
and $a$ an arbitrary traceless $N_c\times N_c$ Hermitian constant matrix.
The net effect is that the $g=0$ functional integral
includes configurations in which
parallel transporters around the time circle,
$U \equiv {\cal P} \,( e^{ig \int_0^\beta dx^0 \> A_0})$,
cover the entire gauge group.
It is the resulting integration over $U$ which implements the projection
onto gauge invariant states.
After suitably fixing a gauge,
the remaining functional integral is a well-defined
Gaussian integral over all degrees of freedom other than $a$, the constant
mode of $A_0$, and a non-Gaussian integral over the single matrix $a$.

The Gaussian integral leads to functional determinants
of the covariance (or small fluctuation) operators defined
by the quadratic part of the action.
For the scalar fields, this operator involves the sum of the
square of a covariant time derivative,
which depends on $a$ and the chemical potentials,
a spatial Laplacian, and the curvature-induced mass term.
The gauge field covariance operator and the square of the fermion operator
have completely analogous forms, but with the relevant Laplacian acting on
vector or spinor fields, respectively, and without any mass term.
Eigenfunctions of these operators are exponentials in time,
$e^{i\omega_k t}$, multiplied by scalar, vector, or spinor
$S^3$-spherical harmonics.
The frequencies $\omega_k$ are quantized Matsubara frequencies, and
equal $2k\pi /\beta$ for bosons and $(2k{+}1)\pi/\beta$ for fermions.
The eigenvalues of the Laplacian on the three-sphere
may be regarded as squares of discrete spatial momenta,
and scale as $1/R^2$, where $R$ is the radius of the sphere.
We will generally set $R=1$, and assume all momenta
and energies are measured in the units of $1/R$.

\begin{TABLE}[ht]
{
\centerline{
\begin{tabular}{l c|c c}
\multicolumn{2}{c|}{field}
& eigenvalue  & degeneracy \\ \hline
transverse vector  &  ${\bf A}_{\perp}$ & $(h+1)^2$  & $2h\,(h+2)$	\\ 
longitudinal vector & $\nabla F$    &  $h\,(h+2)$   & $(h+1)^2$	\\
real scalar & $A_0$   &  $h\,(h+2)$   & $(h+1)^2$	\\
conformal scalar & $\Phi_A$  &  $(h+1)^2$  & $(h+1)^2$	\\
Majorana spinor & $\psi_i$  &  $(h+\half)^2$& $h\,(h+1)$
\end{tabular}}
\caption
    {\footnotesize
    Eigenvalues (in units of $1/R^2$) and degeneracies of the spatial
    part of the small fluctuation operator for scalar, spinor, and
    vector fields.
    The relevant operator is just the Laplacian on $S^3$
    for all fields except conformally coupled scalars,
    where it includes a shift by
    $1/R^2$ due to the curvature induced mass term.
    Representations of the $SO(4)$ isometry group are labeled by
    $h = 0,1,2,\cdots$,
    except for the longitudinal vector $\nabla F$, where $h$ starts from 1.
    }%
\label{table:harmonics}
}
\end{TABLE}

The eigenvalues and associated degeneracies of the
spatial part of the small fluctuation operator, equal to the
spatial Laplacian plus the mass term (for scalars), are shown in 
Table \ref{table:harmonics}.
Representations of the $SO(4)$ isometry group are labeled by
the integer $h$ which runs over all non-negative values;
$h/R$ may be regarded as a discrete spatial momentum.%
\footnote
    {
    For the more familiar case of spherical harmonics $\{Y^l_m\}$ on $S^2$,
    the non-negative integer $l$ labels the $SO(3)$ irreducible 
    representation whose
    Laplacian eigenvalue is $l\,(l+1)$, while the integer $m$ labels
    the basis vectors of the irreducible representation space
    whose dimension is $2l+1$.
    }
The degeneracy factors are the dimension of 
the representations at each value of $h$.
Spatial components
of the gauge fields have been decomposed into a transverse
({\em i.e.}, divergenceless) vector field
${\bf A}_\perp$ and a gradient of a scalar, $\nabla F$.
The longitudinal vector $\nabla F$ vanishes for constant $F$,
and therefore $h$ starts from $1$ in this case. 
The spatial small fluctuation operator
for the conformally coupled scalar fields $\Phi_A$
includes their curvature induced mass term,
so their eigenvalues are $h(h+2)+1=(h+1)^2$.
The time component of the gauge field, $A_0$, is a scalar
with respect to the $SO(4)$ isometry and has no mass.
Note that only $A_0$ has a zero eigenvalue and its associated
eigenfunction is constant on $S^3$.
The lack of zero modes for divergenceless
vector and spinor fields is due to the topology of $S^3$,
which cannot support constant tensors of rank higher than zero.

The complete small fluctuation eigenvalues equal these spatial
eigenvalues plus the square of a Matsubara frequency
shifted by an amount proportional to a difference of eigenvalues of
the matrix $a$ and (for scalars and fermions)
by $i$ times a chemical potential.
Details are given in Appendix \ref{MatrixModelDerivation}.
For the scalar fields $\Phi_A$,
the complete small fluctuation eigenvalues are
\begin {equation}
    (\omega_k + q_m - q_n \pm i \mu_p)^2 + (h+1)^2 \,,
\label {eq:scalar eigs}
\end {equation}
where $\{q_n\}$ are the real eigenvalues of $a$,
$\omega_k = 2\pi k/\beta$,
$m,n = 1, \cdots, \Nc$,
and $h = 0, 1, \cdots$.
Note that if the magnitude of the chemical potential $\mu_p$
exceeds unity (times $1/R$) then, for $m=n$ and $h=k=0$,
the real part of the eigenvalue (\ref {eq:scalar eigs})
becomes negative --- which means that the Gaussian integral
fails to be well-defined.
This can be seen directly from the Lagrangian
(\ref{fullLagrangian}),
which shows that non-zero chemical potentials
act like a negative mass squared for static diagonal components
of the scalar field $\Phi_A$.
If the chemical potential exceeds the $1/R$ curvature induced mass,
then the Euclidean action becomes unbounded below and the partition
function ceases to exist.
So $|\mu_p| = 1/R$ represents a boundary of the phase diagram
of the free theory.
In the remainder of this section we assume all chemical potentials
are less than $1/R$ (in magnitude).

The small fluctuation operator for the
static (zero frequency) component of $A_0$ has no temporal contribution
and is just the scalar Laplacian on $S^3$.
The presence of a zero eigenvalue of the small fluctuation operator
for $A_0$ illustrates the necessity for separating the constant
mode of $A_0$ from the other degrees of freedom,
as done in (\ref {eq:separate}).
The contributions from the non-constant part of $A_0$
and the longitudinal part of the spatial gauge field,
$\nabla F$, end up canceling the contributions from
gauge-fixing ghosts (which are also scalars with respect to $SO(4)$).
The logarithms of the resulting functional determinants
involve a sum over all Matsubara frequencies and a sum
over the discrete momentum $h$ labeling $S^3$ spherical harmonics.
As shown in Appendix \ref{MatrixModelDerivation},
the Matsubara sum may be performed explicitly and the result
for the logarithm of the Gaussian integral may be cast in the form
of an effective action for the single $SU(\Nc)$ matrix 
$U \equiv \exp(i\beta a)$,
\begin{equation}\label{eq:effS}
	S_{\rm eff}(U)
	=
	-\sum_{n=1}^\infty\frac{1}{n}\left\{z_B(x^n)
	+(-1)^{n+1}z_F(x^n)\right\}[\tr (U^n)\tr (U^{\dagger n})-1]
								\;,
\end{equation}
where $x\equiv e^{-\beta}=e^{-1/T}$ and
we have defined the ``single particle'' partition functions:
\begin{subequations}
\label{singlepp}
\begin{align}
	z_B(x)&\equiv z_S(x)+z_V(x)\,,\\[5pt]
	z_V(x)&\equiv \frac{6x^2-2x^3}{(1-x)^3} \,,\\
	z_S(x)&\equiv \frac{x+x^2}{(1-x)^3}
	\left(x^{\mu_1}+x^{-\mu_1}+x^{\mu_2}+x^{-\mu_2}
				+x^{\mu_3}+x^{-\mu_3}\right)\,,\\
	z_F(x)&\equiv \frac{2x^{3/2}}{(1-x)^3} \,
	\bigl(x^{\frac{1}{2}\mu_1}+x^{-\frac{1}{2}\mu_1}\bigr)
	\bigl(x^{\frac{1}{2}\mu_2}+x^{-\frac{1}{2}\mu_2}\bigr)
	\bigl(x^{\frac{1}{2}\mu_3}+x^{-\frac{1}{2}\mu_3}\bigr) \,.
\end{align}
\end{subequations}

To obtain the grand canonical partition function,
we must integrate over the remaining single matrix $a$,
or equivalently over the group element $U$,
\begin{equation}\label{matrixZ}
	Z(x)=\int dU \> \exp[-S_{\rm eff}(U)] \;.
\end{equation}
The required measure $dU$ is Haar measure on the group $SU(\Nc)$.
Even though we have derived the matrix model
(\ref {eq:effS})--(\ref {matrixZ})
specifically for $\Nfour$ $SU(\Nc)$ SYM theory,
the model dependence is only in the specific field content
and group character.
The generalization to other gauge
theories in the zero coupling limit is straightforward.

\subsection{Phase Structure of the Free Theory}\label{phases}

The reduced theory (\ref{matrixZ}) is a single matrix model. 
In the large $\Nc$ limit, this can be solved in a manner
similar to the Gross-Witten model \cite{Gross:1980he}. 
The required analysis is a straightforward generalization of 
the zero chemical potential case discussed in Refs.
\cite{Aharony:2003sx,Sundborg:1999ue}.
Hence we will only sketch the procedure; interested readers should 
refer to Refs.~\cite{Aharony:2003sx,Sundborg:1999ue} for details.%
\footnote
    {
    The analysis in
    Refs.~\cite{Sundborg:1999ue,Aharony:2003sx}
    was carried out with $SU(\Nc)$ and $U(\Nc)$ gauge groups,
    respectively.
    But the difference between the gauge groups is negligible
    in the large $\Nc$ limit,
    and both treatments yield the same
    action (\ref {effaction}).
    }

After rewriting the integration measure in terms of the eigenvalues
of $U$, and introducing the eigenvalue
distribution function, $\rho(\theta)$,
one arrives at the following form of the effective action
for the eigenvalue distribution,%
\footnote
    {%
    The eigenvalue distribution $\rho(\theta)$ must be non-negative
    and satisfy the normalization condition
    $\int_{-\pi}^\pi d\theta \, \rho(\theta) = 1$.
    The positivity condition implies constraints on the Fourier
    coefficients $\{\rho_n\}$ which limit how large any particular
    coefficient can grow.
    Because of this boundary on the space of allowable $\rho_n$,
    the effective action (\ref {effaction}) remains bounded below,
    with a well-defined minimum, even when the coefficients $V_n$
    are not all positive.
    This boundary is irrelevant
    in the disordered phase where all $V_n$ are positive
    and the minimum of $S_{\rm eff}$ lies at $\rho_n = 0$ (for $n > 0$).
    When one or more of the $V_n$ are negative, then the minimum
    lies on the boundary and its presence is essential.
    }
\begin{equation}\label{effaction}
    S_{\rm eff}[\rho]=\Nc^2\sum_{n=1}^\infty \> V_n \, \rho_n^2
								\;,
\end{equation}
where
$\rho_n\equiv \int_{-\pi}^{\pi}d\theta\,\rho(\theta)\cos(n\theta)$ 
are the Fourier series coefficients of the eigenvalue distribution,
and%
\footnote
    {
    The $1/n$ term in $V_n$ comes from
    the Haar measure $dU$.
    Expressed in terms of the eigenvalues of the matrix $U$,
    this generates a Van der Monde determinant which may be written
    as a contribution of
    $
	-\Nc^2\int d\theta \, d\theta' \> \rho(\theta) \, \rho(\theta') \,
	\ln |2\sin(\frac\theta 2{-}\frac{\theta'}2)|
    $
    to $S_{\rm eff}[\rho]$.
    Inserting the identity
    $\ln |2\sin \frac x2| =-\sum_{n=1}^{\infty}\frac 1n \cos nx$,
    and assuming that $\rho(\theta)$ is an even function of $\theta$,
    leads to the form
    $\Nc^2 \sum_{n=1}^\infty \frac 1n \, \rho_n^2$.
    }
\begin{equation}\label{Vn}
	V_n\equiv \frac{1}{n}\left\{1-
	\left[ z_B(x^n)+(-1)^{n+1}\, z_F(x^n) \right]
	\right\}
								\;.
\end{equation}
The factor of $\Nc^2$ multiplying the effective action (\ref {effaction})
implies that the minimum value of this effective action determines
the leading large $\Nc$ behavior of the free energy
$F \equiv -\beta^{-1} \ln Z$,
with fluctuations in the eigenvalue density only generating
subleading $\mathcal O(\Nc^0)$ contributions to the free energy.
As $\Nc \to \infty$,
\begin {equation}
    F
    =
    \mathop{\min}\limits_{\{\rho\}} \> \frac {S_{\rm eff}[\rho]}\beta
    +
    \mathcal O(\Nc^0)
    \,.
\end {equation}

In the effective action (\ref {effaction})
each positive $V_n$ acts as a repulsive potential
for the eigenvalue distribution,
while negative $V_n$'s act as attractive
potentials. Thus the condition
\begin{equation}
	z_B(x^n)+(-1)^{n+1}z_F(x^n)<1
\label {eq:zineq}
\end{equation}
ensures that all $V_n$ are positive and
implies that the system is in the phase where eigenvalues repel
and the uniform distribution $\rho(\theta) = \frac 1{2\pi}$
characterizes the equilibrium state.
As long as all $|\mu_p|$ are less than one,
which is required for the existence of the grand canonical ensemble,
it is easy to see that our modified single particle 
partition functions (\ref{singlepp}) increase monotonically with $x$.
Hence, the $n=1$ term in 
the inequality (\ref {eq:zineq}) gives the
most stringent condition.
Therefore,
the curve in the $\mu$-$T$ diagram
determined by the threshold condition
\begin{equation}\label{hagline}
	z_B(x)+z_F(x)=1
\end{equation}
gives the boundary of the phase in which the eigenvalue
distribution is uniform and $\tr(U^n)$ has vanishing expectation value
(for all $n \ne 0$).

In the free-field and large $\Nc$ limit,
the Polyakov loop expectation value is just $\rho_1$:
\begin{equation}
    \left\langle 
    \coeff 1\Nc \tr \, {\cal P}\left( e^{ig \int_0^\beta dx^0 \> A_0}\right)
    \right\rangle
    = \left\langle \coeff 1\Nc\tr e^{i\beta a} \right\rangle
    = \left\langle \coeff 1\Nc\tr U \right\rangle
    = \int_{-\pi}^\pi d\theta \> \rho(\theta) \, e^{i\theta}
    = \rho_1 \,.
\end{equation}
So in the disordered phase, where
the eigenvalues of $U$ are uniformly distributed on the unit circle,
the Polyakov loop expectation vanishes.
In the ordered phase, where the eigenvalue distribution is non-uniform,
the Polyakov loop will have a non-zero expectation value.
The Polyakov loop transforms non-trivially under the $\mathbb{Z}_{\Nc}$
center of the gauge group, and its expectation value is an order parameter
for the realization of this symmetry.
Thus the $\mathbb{Z}_{\Nc}$ symmetry is unbroken in the
disordered phase, and spontaneously broken in the ordered phase.
As emphasized in the Introduction,
it is this behavior of the Polyakov loop and the associated realization
of the $\mathbb{Z}_{\Nc}$-symmetry which motivates calling the
disordered phase ``confining'' and the ordered phase ``deconfined''.%
\footnote
    {
    The expectation value of the Polyakov loop may be interpreted as
    $\exp[-\beta\Delta F]$ where $\Delta F$ is the free energy difference
    between equilibrium states in which one has, or has not, added a
    fundamental representation static test quark.
    In the confining phase this free energy difference is infinite
    and the expectation value of the Polyakov loop vanishes,
    while in the deconfined phase the free energy difference is finite
    and the expectation value is non-zero.
    Hence the Polyakov loop may be viewed as a confinement/deconfinement
    order parameter.
    But strictly speaking, the Polyakov loop expectation value
    is {\em defined} via cluster decomposition from the large distance limit
    of the Polyakov loop two-point function.
    It is the two-point function which, in infinite volume,
    embodies the operational
    definition of confinement in terms of the free energy needed
    to separate a test quark and antiquark to infinity.
    }

\begin{FIGURE}
{
\centerline{\scalebox{.61}{\includegraphics{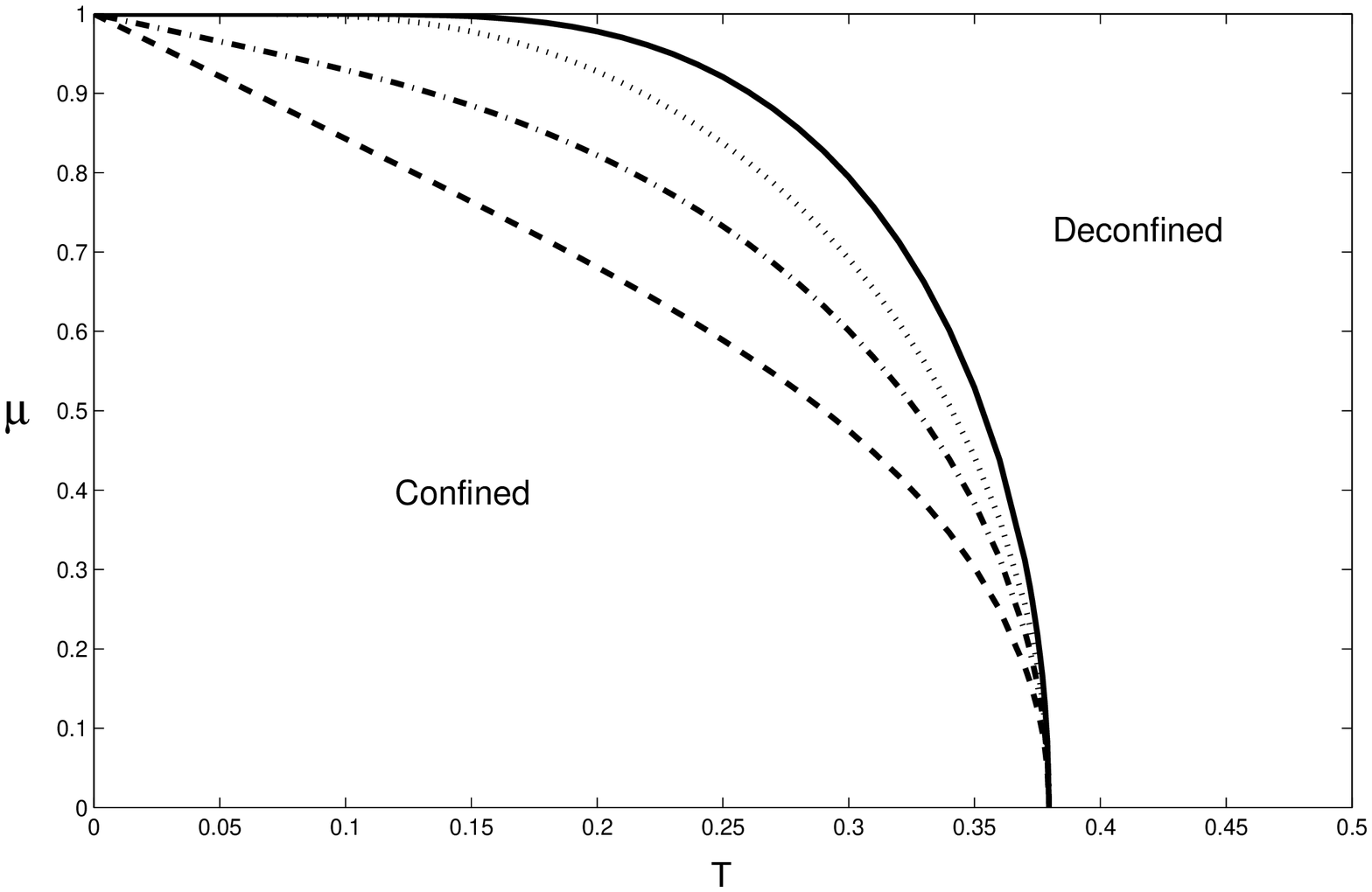}}}
\caption{\footnotesize
The confinement/deconfinement phase transition line in the $\mu$-$T$ plane
for the free theory.
Both $\mu$ and $T$ are measured in the units of $1/R$.
The values of $(\mu_1,\mu_2,\mu_3)$ for each line are as follows.
Solid line: $(\mu,0,0)$,
dotted: $(\mu,\mu/2,\mu/2)$,
dash-dotted: $(\mu,\mu,0)$,
dashed: $(\mu,\mu,\mu)$.
}\label{hagphase}%
}
\end{FIGURE}

Returning to equation (\ref{hagline}), it is straightforward
to plot where, in the $\mu$-$T$ plane, solutions to this condition lie.
Fig.~\ref{hagphase} shows the resulting curves
in four representative cases where
$(\mu_1,\mu_2,\mu_3)=(\mu,0,0)$, $(\mu,\mu,0)$, 
$(\mu,\mu,\mu)$, or $(\mu,\mu/2,\mu/2)$.
Since there are three independent chemical potentials, the full
phase diagram is four dimensional, but the slices shown in 
Fig.~\ref{hagphase} illustrate the general behavior.
When all chemical potentials are zero,
the phase transition line reaches the $T$-axis at $T\approx0.38/R$
agreeing, as it should, with Ref.~\cite{Aharony:2003sx}.
If $|\mu|$ denotes the maximum magnitude of the three chemical potentials
then, as discussed earlier, $|\mu| = 1/R$ is a boundary of the
free field phase diagram.
In the vicinity of zero temperature, the relevant terms in Eq.~(\ref{hagline})
are
\begin{equation}
	e^{-\beta(1-\mu_1)}
	+
	e^{-\beta(1-\mu_2)}
	+
	e^{-\beta(1-\mu_3)}
	+
	2\, e^{-\frac 12 \beta(3-\mu_1-\mu_2-\mu_3)}
	=
	1
							\;,
\end{equation}
where the first three exponentials are from $z_S(x)$,
the last one is from $z_F(x)$, and the radius of the sphere $R$ is set to
unity.
From this expression, one sees that the maximal chemical potential must
approach $1$ as $T\rightarrow 0$.
Thus, the phase transition line
necessarily ends at the boundary $|\mu|=1$ when the temperature
falls to zero. The slope of the transition line at zero temperature
depends on how many chemical potentials approach unity.
It is easy to see that
the limiting slope of the transition line is zero
if a single chemical potential is turned on.
The limiting slope is $-\ln 2$ when two equal chemical potentials are
turned on,
and $-\ln 4$ with three equal potentials.

\subsection{Order of the Phase Transition}\label{1storder}

In the confining phase, the density of eigenvalues of the
matrix $U$ is constant and
all (non-trivial) Fourier coefficients of $\rho(\theta)$ vanish.
The minimum value of the effective action (\ref {effaction}) vanishes,
and hence the free energy in this phase is
$\mathcal{O}(\Nc^0)$, not $\mathcal{O}(\Nc^2)$, as $\Nc \to \infty$.
In the deconfined phase there is a non-constant density of eigenvalues,
the minimum value of the effective action (\ref {effaction})
is non-zero, and the free energy is $\mathcal O(\Nc^2)$.
In other words, $\lim_{\Nc\to\infty} F/\Nc^2$ is non-zero in the
deconfined phase but vanishes identically in the confined phase.
The free energy must be continuous across any phase transition
(but its derivatives need not be),
so the coefficient of the $\mathcal O(\Nc^2)$ part of the free energy
must vanish as one approaches the phase transition line
from the deconfined side.
If it vanishes linearly (with temperature or chemical potential)
then first derivatives of the free energy will be discontinuous
and the $\Nc=\infty$  phase transition is first order.
If the $\mathcal O(\Nc^2)$ free energy vanishes faster than linearly
as the transition line is approached,
then the $\Nc=\infty$ phase transition is continuous.

The phase transition line is determined by the condition
$
	z_B(x)+z_F(x)=1
$,
and the left-hand side is a monotonically increasing
function of temperature.
Therefore as we approach from the deconfined side,
we can analyze the local behavior near the phase transition
line by expanding the solution to the
matrix model in powers of $\epsilon^2 \equiv z_B(x)+z_F(x)-1$,
with $\epsilon$ real.
Such an expansion was carried out in Ref.~\cite{Aharony:2003sx}
for the case of zero chemical potentials, with the result that
(for $\epsilon^2 > 0$)
\begin{equation}\label{expansion}
	\lim_{\Nc\to\infty} \>
	\frac{\beta F}{\Nc^2}=-\frac{\epsilon^2}{4}+\mathcal{O}(\epsilon^3) \;.
\end{equation}
(Additional higher order terms are
also obtained in Ref.~\cite{Aharony:2003sx}.)
The analysis is independent of whether the single particle partition 
functions contain chemical potentials,
so the above result is equally valid in our case with
chemical potentials.

To see if the phase transition is first order, one must merely
determine how $\epsilon^2$ in the result (\ref {expansion})
depends on $T-T_c$ or $\mu - \mu_c$.
The function $\epsilon^2 \equiv z_B(x)+z_F(x)-1$ depends
analytically on $T$ and $\mu$ and there is no reason
for its derivatives with respect to $T$ or $\mu$ to vanish
on the line where $\epsilon^2$ crosses zero.
It is straightforward to check numerically
that this is, in fact, the case.
One finds that $\epsilon^2$ vanishes linearly with $T-T_c$ and $\mu-\mu_c$
as one approaches a point $(T_c,\mu_c)$ on the
phase transition line from the deconfined phase.%
\footnote
    {
    For example, with only one chemical potential turned on,
    $(T_c,\mu_c)=(0.35,0.53)$ (measured in the units of $1/R$)
    is a point on the phase transition line
    and
    $
	    \frac{\beta F}{\Nc^2}\sim -2.4\, (T-T_c) - 0.31\, (\mu-\mu_c)
    $.
    When all three chemical potentials are equal,
    $(T_c,\mu_c)=(0.35,0.32)$ lies on the transition line and
    $
	    \frac{\beta F}{\Nc^2}\sim -2.5\, (T-T_c) - 0.48\, (\mu-\mu_c)
    $.
    }
Hence the rescaled large $\Nc$ free energy,
$\lim_{\Nc\to\infty} F/\Nc^2$,
has a discontinuous first derivative as one crosses the phase transition line,
showing that the large $\Nc$ confinement/deconfinement phase transition,
at zero coupling, is first order.

First order phase transitions are normally robust phenomena with regard
to perturbations, such as a change in the coupling
in the underlying theory.
So one might expect that a first order confinement/deconfinement
transition in the zero coupling limit of the theory would
imply that the transition must remain first order for some
non-zero range of couplings.
In the case at hand, however, the situation is more subtle.

At a first order transition,
there are multiple equilibrium states so that, as the
transition is approached, the limit depends on the direction of the approach.
At a typical first order transition, the coexisting
equilibrium states are separated by free energy barriers,
and it is the presence of these free energy barriers which
lead to characteristic phenomena associated with first order transitions
such as superheating or supercooling.
But in $\Nc=\infty$ gauge theories at zero coupling,
such as the $\Nfour$ SYM theory under discussion,
the free energy at phase coexistence does not have isolated minima,
but rather has a flat direction.
This may be seen directly in the effective action (\ref {effaction})
for the density of eigenvalues.
The coefficient $V_1$ vanishes at the deconfinement transition,
so minimizing the effective action leaves the Fourier coefficient
$\rho_1$ completely undetermined.
Approaching the phase transition line
from the confined or deconfined phases leads, at the transition,
to the equilibrium states with minimal or maximal values of $\rho_1$,
respectively.
But these states are not separated by any free energy barrier.%
\footnote
    {%
    Because of this, some might quarrel with calling this
    a first order transition, even through the free energy
    shows a kink (in temperature or chemical potential)
    as one crosses the transition line.
    Following the analysis of Ref.~\cite{Aharony:2003sx},
    one may show that if the transition line is approached from
    the deconfined phase, then the Polyakov loop expectation value $\rho_1$
    has a limiting value of $1/2$,
    independent of the chemical potential.
    }

The lack of a free energy barrier separating the coexisting
equilibrium states (in the zero coupling theory) means that
a small perturbation, such as turning on an
infinitesimal non-zero coupling,
can have a large effect.
A non-zero coupling can lift the flat direction
and produce either a first order transition (with a barrier
separating coexisting states) or a second order transition,
depending on the sign of the $(\rho_1)^4$ term which is
induced in the effective action for the density of eigenvalues.
As discussed in greater detail in Ref.~\cite{Aharony:2003sx},
a three-loop calculation is needed to determine this sign.
The required three-loop calculation has not yet been done
(even for zero chemical potential), so the
order of the confinement/deconfinement transition
in $\Nfour$ super Yang-Mills theory at small but non-zero
coupling is currently unknown.%
\footnote
    {
    For pure Yang-Mills theory on a sphere,
    the corresponding three-loop calculation has been performed
    \cite{Aharony:2005bq}, and in this case
    the large $\Nc$ confinement transition remains first order
    at small but non-zero coupling.
    }

\subsection {Canonical vs. Grand Canonical}\label{canonical}

To understand how our results compare with the picture of
Ref.~\cite{Basu:2005pj},
one must convert from the grand canonical to the canonical ensemble
(or vice-versa).
Starting from the grand canonical free energy
$F(\beta,\mu_p) \equiv -\beta^{-1} \ln Z(\beta,\mu_p)$,
the $R$-charges are given by
$Q_p = -\frac {\partial F}{\partial \mu_p} $,
and the thermodynamic potential of the canonical ensemble is
$A(\beta,Q_p) \equiv F + \sum_p \mu_p \, Q_p$
(with the chemical potentials now viewed as functions of the charges).
Given the canonical ensemble potential as function of temperature and charge,
the chemical potentials are given by
$\mu_p = \frac {\partial A}{\partial Q_p}\,$.

Within the deconfined phase, the free energy and the $R$-charges
are both of order $\Nc^2$, so the rescaled charges
\begin {equation}
    q_p \equiv \lim_{\Nc\to\infty} \frac{Q_p}{\Nc^2}
\end {equation}
are appropriate observables in the large $\Nc$ limit.
Ref.~\cite{Basu:2005pj} considered thermodynamics with
a fixed non-zero value for the total rescaled charge $q \equiv \sum_p q_p$.

Consider, for simplicity, the case of equal chemical potentials.
In the interior of the deconfined phase, there is a one-to-one
mapping between the chemical potential $\mu$ and the rescaled charge $q$.
But within the confined phase,
the $R$-charges (as well as the free energy)
are $\mathcal O(1)$, so the rescaled charge $q$ vanishes identically.
At any fixed temperature $T$
below  the $\mu = 0$ confinement transition temperature,
if one increases the chemical potential starting from zero
then $q$ remains identically zero until one reaches the 
confinement/deconfinement transition at $\mu = \mu_c(T)$.
Since the transition (in the free theory) is first order,
the rescaled charge $q$ jumps discontinuously across the transition
to some non-zero value $q_c(T)$,
which is the minimal value of $q$ (at the given temperature)
within the deconfined phase.
As the chemical potential is further increased, the charge
continues to grow, eventually diverging at the edge of the phase diagram
({\em i.e.}, as $\mu \to 1$).

From this description, it might seem that it is impossible to obtain
an equilibrium state with $0 < q < q_c(T)$ --- but this is wrong.
The essential point is that first order transitions always involve
phase coexistence.
The $q = q_c(T)$ and $q = 0$ statistical ensembles produced 
by taking $\mu \to \mu_c(T)$ from above and below, respectively,
in the grand canonical ensemble are extremal equilibrium states.
But any statistical mixture of these two states is also a valid
equilibrium state in the $\Nc \to \infty$ limit.

Therefore, the canonical description of the phase diagram
shown in Fig.~\ref{hagphase}
will have a deconfined phase for $q > q_c(T)$ in which
the susceptibility
$
    \frac {\partial\mu}{\partial q}
    = \frac 1{\Nc^2} \, \frac{\partial^2 A}{\partial q^2}
$ 
is positive
(so the chemical potential increases monotonically with $q$),
together with a phase coexistence region for $0 < q < q_c(T)$
in which the the chemical potential
$
    \mu = \lim_{\Nc\to\infty} \frac 1{\Nc^2} \, \frac {\partial A}{\partial q}
$
is independent of $q$.
Within the phase coexistence region the
(rescaled, large $\Nc$ limit of the) thermodynamic potential,
$\lim_{\Nc\to\infty} A/\Nc^2$, is simply equal to $\mu(T) \, q$.
Only at $q = 0$ will one have the pure confined phase.
This is illustrated in Fig.~\ref{fig:canonical}

\begin{FIGURE}[t]
{
\centerline{\scalebox{.45}{\includegraphics{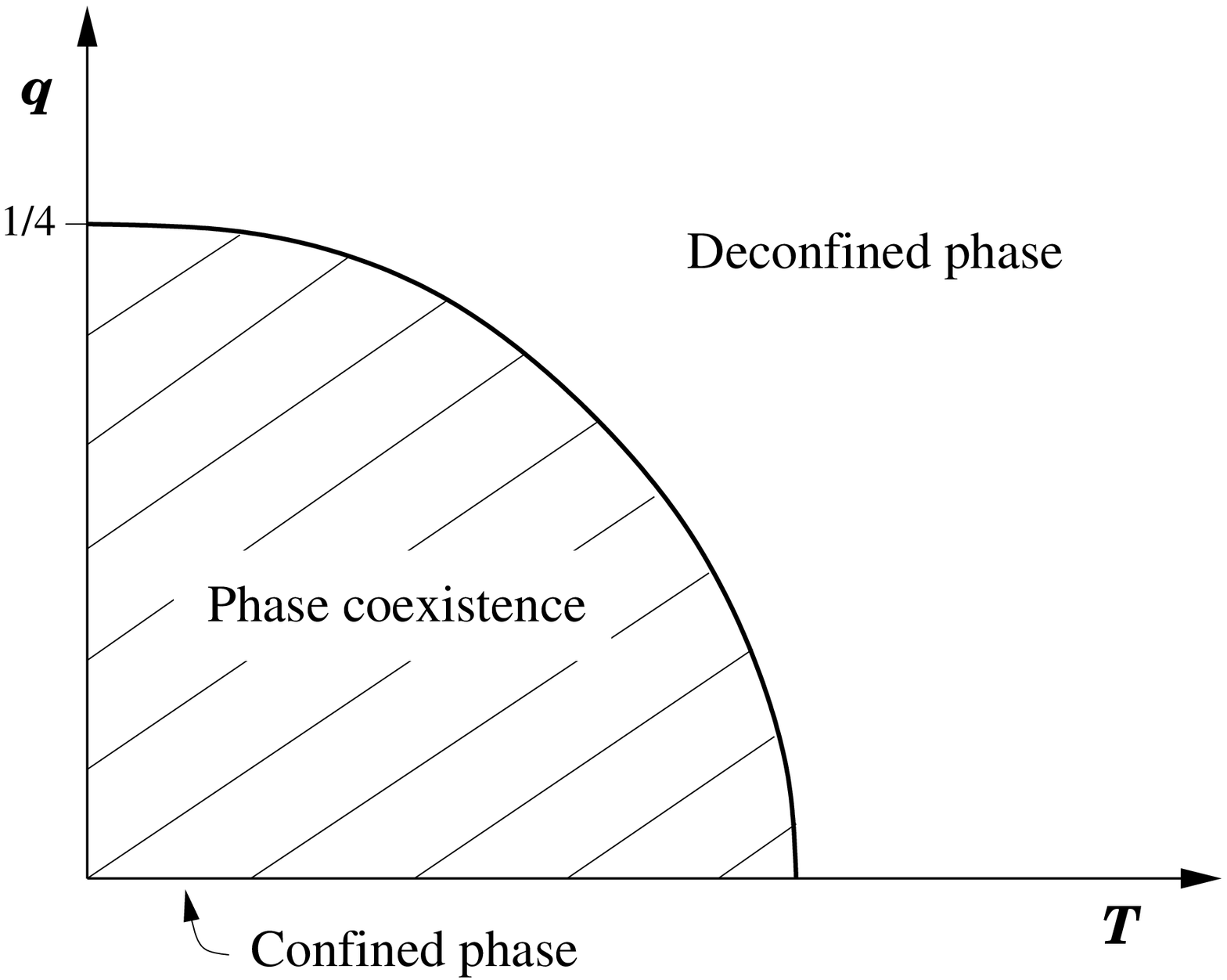}}}
\vspace*{-1em}
\caption{\footnotesize
Schematic phase diagram for the free theory in the $q$-$T$ plane.
Within the phase coexistence region, the chemical potential
is independent of the charge $q$, and equal to its value $\mu_c(T)$ on the
confinement/deconfinement transition line.
Inside the deconfined phase, the chemical potential (at fixed $T$)
increases with increasing $q$.
Only the $T < T_c$ segment of the $q = 0$ axis corresponds to
the pure deconfined phase.
The maximal charge $q_c^{\rm max} = 1/4$,
and the maximum transition temperature $T_c^{\rm max} \approx 0.38/R$.
}\label{fig:canonical}%
}
\end{FIGURE}

Although the authors of Ref.~\cite{Basu:2005pj} did not explicitly evalute 
the chemical potential produced by their thermpdynamic potential,
or notice its independence on $q$ for their ``small $q$'' extremum,
the results of Ref.~\cite{Basu:2005pj} are completely consistent
with the above description of phase coexistence in the $q$-$T$
phase diagram.%
\footnote
    {
    Within the truncation used in Ref.~\cite{Basu:2005pj},
    the thermodynamic potential in the confined phase is given by
    $
	\beta A(\beta,q)
	=
	\min_{\rho} \left[ \Nc^2 \, \rho^2 - S_{\rm eff}(\rho) \right]
    $,
    where $S_{\rm eff}(\rho)$ is their Eq.~(3.19).
    Evaluating this gives a result proportional to $q$,
    and hence a chemical potential independent of $q$.
    These authors also constructed a phenomenological model based
    on a deformation of the effective action which can mimic the effect
    of turning on a weak gauge coupling.
    For this generalization,
    assuming the transition remains first order, one again finds that
    $A$ is linear
    and the chemical potential is independent of $q$
    in the region of the phase diagram where the thermodynamic potential
    is minimized by the stable small $q$ solution
    (denoted solution $I$ in \cite{Basu:2005pj}).
    Hence this region again corresponds to phase coexistence,
    and the phase diagram resulting from this model resembles
    Fig.~\ref{fig:canonical} above.
    }

\section{Weak Coupling, High Temperature Effective Theory}\label{hightemp}

We now
consider the theory in the high temperature regime, $T \gg 1/R$,
with a small non-zero coupling.
The relevant expansion parameter will be $1/(TR)^2$, so the
high temperature regime may also be viewed as the regime
of large spatial radius at a fixed temperature.
Consequently,
for observables which are primarily sensitive to the scale $T$,
the curvature of the $S^3$ will produce only
small corrections to flat space results.

Let $\lambda \equiv g^2\Nc$ denote the usual 't Hooft coupling,
which is held fixed as $\Nc \to \infty$.
Interesting phenomena in the phase diagram will be found to occur
when $1/(TR)^2$ and $\mu^2/T^2$ are both of order $\lambda$,
so we will focus on this region of parameter space.

At non-zero temperature,
one may always decompose fields
into Fourier series running over all Matsubara frequencies,
and thereby view every four-dimensional field as an infinite
tower of three dimensional fields.
At high temperature, where
the $1/T$ circumference of the periodic time circle
is small compared to all other scales,
the non-zero Matsubara frequency modes behave
(with respect to physics on length scales large compared to $1/T$)
like fields describing very heavy excitations with $\mathcal{O}(T)$ mass.
As $T\to\infty$,
the non-static heavy modes decouple \cite{Appelquist:1974tg}
and one may integrate them out leaving only the light zero frequency modes.
In this way, one obtains a three dimensional effective theory of the static
modes \cite{Gross:1980br}.
Convenient techniques for systematically constructing this effective theory,
using dimensional continuation to regulate both long and short distance
behavior, were described by Braaten and Nieto
\cite{Braaten:1995cm,Braaten:1995jr}.

There are three scales which are relevant for equilibrium thermodynamics
in weakly coupled high temperature non-Abelian gauge theories
at zero chemical potential:
the temperature $T$\!,
the ``electric mass'' (or inverse Debye screening length) which is
$\mathcal{O}(\!\sqrt\lambda T)$,
and the ``magnetic mass'' (or inverse correlation length of the
static gauge field) which is $\mathcal{O}(\lambda T)$.
One may construct a dimensionally reduced ``electric'' effective
theory, valid on length scales large compared to $1/T$,
by integrating out non-static fluctuations.
One may then construct a further reduced ``magnetic'' effective
theory, valid on length scales large compared to $(\sqrt\lambda \, T)^{-1}$,
by also integrating out static fluctuations on the Debye screening scale.
For QCD, these two effective theories are commonly referred to
as EQCD$_3$ and MQCD$_3$, respectively \cite{Braaten:1995jr}.

For our case of $\Nfour$ super-Yang-Mills theory with non-zero chemical
potentials, it is the ``electric'' effective theory which will
be of interest.
This effective theory,
which we will call ``ESYM$_3$'',
will contain a three dimensional gauge field $\bf A$,
the static component of $A_0$, which will appear as an adjoint scalar field,
and the static components of the original scalars $\Phi_A$.
The fermions, having no zero-frequency components due to their
antiperiodicity, will be completely integrated out.
Operationally, one builds the effective theory by considering
all local operators
which may be constructed from these fields and are
consistent with the symmetries of the theory.
Up to any given order in perturbation theory,
only a finite number of operators with sufficiently low dimensions
are needed.
The required coefficients
are determined by matching the results for a minimal set of observables
which may be evaluated in both the effective and underlying theories
\cite{Braaten:1995cm,Braaten:1995jr}.

The operators in ESYM$_3$ include the identity operator,
the usual gauge invariant derivative terms,
quadratic mass terms for scalar fields,
and scalar interaction terms.
It is convenient to use rescaled fields in the effective theory,
\begin {equation} \label{eq:rescale}
    A_i \equiv \sqrt {Z_1 T} \, \tilde A_i \,,\qquad
    A_0 \equiv \sqrt {Z_2 T} \, \tilde A_0 \,,\qquad
    \Phi_A \equiv \sqrt {Z_3 T} \> \tilde \Phi_A \,,
\end {equation}
with $Z_1$, $Z_2$, and $Z_3$ dimensionless wavefunction renormalization factors,
and $\tilde A_\mu$ and $\tilde \Phi_A$ now having canonical dimension 1/2.
The wavefunction renormalization factors may be chosen so that
the Lagrangian density of the effective theory has the form
\begin {equation}\label{LESYM}
    {\cal L}_{{\rm ESYM}_3}
    =
    f + \,\tr\! \left[
	    \coeff{1}{2} (\tilde F_{ij})^2
	    + (D_i \tilde A_0)^2 + M_D^2 \, \tilde A_0^2
	    + (D_i\tilde \Phi_A)^2 + m_A^2 \, \tilde \Phi_A^2
	    \right]
	    + {V}(\tilde A_0,\tilde \Phi_A) \,,
	    \vphantom{\Bigg|}
\end{equation}
where $D_i \equiv \partial_i + i g_3[\tilde A_i,\cdot]$
and $\tilde F_{ij} = [D_i,D_j]$, with $g_3 \equiv g \sqrt T$
the dimensionful gauge coupling appropriate for a three-dimensional theory.

The coefficient of the identity operator, $f$, is the 3-$d$ effective theory
version of a cosmological constant,
and represents the contribution to the free energy density from modes
which have been integrated out.
Because we are working at high temperature (or large volume)
and only integrating out heavy modes with short $\mathcal O(1/T)$
correlation lengths, the computation of the free energy density
due to heavy modes may be performed entirely in flat space, treating
the curvature and chemical potential induced scalar mass terms
as small perturbations.
Corrections to this approximation will vanish exponentially fast in $TR$
(or faster than any power of $\lambda$, since we are assuming that
$TR$ is of order $1/\sqrt\lambda$).
Details of the calculation are described in appendix \ref {IDOp}.
The result, through order $\lambda$, is
\begin{equation}\label{coeffidentity}
	f=-\frac{\Nc^2\,T^3}{12}\bigg\{2\pi^2
	-3\lambda
	-\frac{3}{T^2R^2}
	+ \sum_{p=1}^3 \> \frac {2\mu_p^2}{T^2}
	+ \sum_{i=1}^4 \> \frac {\tilde\mu_i^2}{T^2} 
	+\mathcal{O}(\lambda^2)\bigg\} \,.
\end{equation}

The mass parameters $m_A^2$ of the scalar fields $\tilde \Phi_A$
receive tree-level contributions from the curvature coupling and the
chemical potentials, plus $\mathcal O(\sqrt\lambda \, T)$
one-loop thermal corrections, so that
\begin{equation}\label{scalarmass}
    m_A^2 = R^{-2} - \mu_A^2 + \delta m^2(T) \,,
\end{equation}
where we introduce (for later convenience)
\begin {equation}
    \mu_A \equiv \begin{cases} \mu_1 \,, & A = 1 \mbox { or } 2; \cr
			       \mu_2 \,, & A = 3 \mbox { or } 4; \cr
			       \mu_3 \,, & A = 5 \mbox { or } 6.
		    \end{cases}
\end {equation}
Because we will be interested in the regime where the tree-level contributions
nearly cancel, including the thermal mass correction $\delta m^2(T)$
is essential.
This contribution,
as well as the one-loop scalar wavefunction renormalization
(which will also be needed)
may be obtained by matching the two-point scalar correlator
in the original theory at zero frequency and low spatial momentum
with the corresponding correlator in
the effective theory.
The computation is shown in Appendix \ref{ScalarMass}.
The result for the thermal mass (also derived in 
Refs.~\cite{Kim:1999sg,Fotopoulos:1998es,Vazquez-Mozo:1999ic})
is
\begin {equation}\label{MassCorrection}
    \delta m^2(T) = T^2 \left[ \, \lambda + \mathcal O(\lambda^2) \right] .
\end {equation}
One-loop fluctuations on the scale $T$ also produce an
$\mathcal O(\sqrt\lambda\, T)$ mass for $\tilde A_0$.
The resulting Debye mass parameter is \cite{Kim:1999sg}
\begin{equation}\label{eq:M_D}
    M_D^2 = T^2 \left[ \, 2 \lambda + \mathcal O(\lambda^2) \right].
\end{equation}

The interaction potential $V(\tilde A_0,\tilde \Phi_A)$
contains both tree-level and fluctuation induced thermal contributions,
\begin{equation}
    V(\tilde A_0,\tilde \Phi_A)
    =
    V_{\rm tree}(\tilde A_0,\tilde \Phi_A)
    +
    \delta V(\tilde A_0,\tilde \Phi_A) \,.
\end{equation}
The tree-level part equals the scalar interaction terms of the
original four-dimensional action, re-expressed in terms of the
rescaled fields (\ref {eq:rescale}) (with the wavefunction renormalization
factors equal to one at lowest order),
\begin{equation}\label{eq:Vtree}
    V_{\rm tree}(\tilde A_0,\tilde \Phi_A)
    =\tr\left\{
    2 g_3 \, \mu_p \, ( [\tilde A_0, \tilde X_p] \, \tilde Y_p )
    + g_3^2 \, (i[\tilde A_0, \tilde \Phi_A])^2
    + \half g_3^2 \, (i[\tilde\Phi_A,\tilde\Phi_B])^2  \right\}
								\,.
\end{equation}
It is straightforward to obtain one-loop contributions to
$\delta V$ that are quartic in $\Phi_A$.
The coefficient of the quartic term can be computed by
evaluating the one-loop heavy-mode contribution to the four-point correlator
of the scalar fields
with vanishing external momenta.
Once again, because we are interested in effects due to
heavy modes with correlation lengths small compared to $R$,
these contributions may be evaluated in flat space,
ignoring both curvature and chemical potential corrections.
The required calculation is performed
in appendix \ref{QuarticSubsection}, with the result
\begin{equation}\label{ESYMquartic}
	\delta V_{\rm quartic}(\tilde\Phi_A)
	=
	\frac{\ln 2}{2\pi^2} \> \frac{g_3^4}{T} \,
	\tr (
	\tilde\Phi^{\rm adj}_B\, \tilde\Phi^{\rm adj}_C \,
	\tilde\Phi^{\rm adj}_B\, \tilde\Phi^{\rm adj}_C ) \,,
\end{equation}
where the trace is in the adjoint representation,
with $(\tilde\Phi^{\rm adj}_A)_{ac} \equiv i \tilde\Phi_A^b f^{abc}$.
Note that no terms cubic in $\tilde\Phi_A$ can be generated,
since they would not respect the $U(1)^3$ subgroup of
the $SU(4)$ $R$-symmetry, which is preserved
even in the presence of chemical potentials.

One may, in principle, continue in a similar manner
and obtain the coefficients of arbitrarily
higher dimensional operators in the effective action.
However, in the special case
where the scalar fields take values in flat directions
(which minimize the tree potential),
one may employ a more elegant background field method.
This method not only confirms the coefficients of the results
(\ref{MassCorrection}) and (\ref {ESYMquartic}),
it also yields all higher order terms, in powers of $\Phi$,
in the one-loop contribution to $\delta V(\Phi_A)$.
This is discussed in Appendix~\ref{BGField} and the result is
\begin{align}
	\delta V_{\text{flat}}(\tilde\Phi_A)
	&=
	\half \pi^2 \> T^3 \, \tr\!
	\bigg[\,
	  (\ln 2)\Big(
		\frac{g_3^2}{\pi^2T^2}
		\sum_A \tilde\Phi_A^\text{adj} \tilde\Phi_A^\text{adj}
		\Big)^2
\nonumber\\
	  &+\sum_{l=3}^{\infty} \> {8}(1-4^{-l+2}) \,
	  \frac{(2l{-}5)!!}{(2l)!!} \, \zeta (2l{-}3)
	  \Big(-\frac{g_3^2}{\pi^2T^2}
		\sum_A \tilde\Phi_A^\text{adj}\tilde\Phi_A^\text{adj}
	  \Big)^{l} \,
	\bigg]
	\;.
\label{eq:Vflat}
\end{align}
These adjoint representation traces
may be written more explicitly in terms
of the eigenvalues $\{ \tilde\lambda_A^m \}$ of $\tilde\Phi_A$
(viewed as an $N_c \times N_c$ matrix).
It will prove convenient to introduce
dimensionless rescaled eigenvalues,
\begin{equation}
    \rho_A^m \equiv \frac{g_3}{\pi T} \> \tilde\lambda_A^m \,,
\label{eq:rho}
\end{equation}
in terms of which
\begin{equation}
    \tr \biggl[\Big(
	    \frac{g_3^2}{\pi^2T^2}
	    \sum_A \tilde\Phi_A^\text{adj} \, \tilde\Phi_A^\text{adj}
	\Big)^l\,\biggr]
    =
    \sum_{m,n} \sum_A \>
    ( \rho_A^m - \rho_A^n )^{2l} \,.
\end{equation}
Note that when the rescaled eigenvalues $\{ \rho_A^m \}$
(and their differences) are parametrically of order one, then
all terms in the series (\ref {eq:Vflat}) are comparable in size.
The potential (\ref {eq:Vflat}) will play an essential role in the
next section, where its behavior will be examined in more detail.

\section{High Temperature Thermodynamics}\label{ESYM3thermodynamics}

\subsection{Deconfined Plasma Phase}\label{sec:symm}

The effective scalar masses
$m_A^2$
[given by Eqs.~(\ref{scalarmass}) and (\ref {MassCorrection})]
are positive as long as the chemical potentials
are sufficiently small,
\begin {equation}
    \mu_A^2 < R^{-2} + \lambda \, T^2 \,.
\label{eq:mumax}
\end {equation}
Assuming this is the case,
the trivial configuration of vanishing scalar fields,
\begin {equation}
    \tilde\Phi_A = 0 \,,\qquad \tilde A_0 = 0 \,,
\label{eq:trivial}
\end {equation}
is a local minimum of the effective potential.
The conditions under which this is the global minimum
will be examined in the next subsection.
Because of the non-zero Debye mass $M_D^2$ [given by Eq.~(\ref {eq:M_D})],
the static mode $\tilde A_0$ has perturbatively small fluctuations in
the weak coupling, high temperature regime.
Hence, the Polyakov loop expectation value
$\langle \tr U(\vec x) \rangle = \langle \tr e^{ i \beta g A_0(\vec x) } \rangle$
is non-zero.%
\footnote
    {
    Explicitly,
    $
    \langle \tr U \rangle 
    = N_c\big[
	1 + \big(1{-}\frac{1}{N_c^2}\big) \frac{\lambda^{3/2}}{8\sqrt{2}\pi}
	+\mathcal{O}(\lambda^2)
      \big]
    $
    in the deconfined plasma phase.
    This follows from the pure Yang-Mills result
    \cite{Gava:1981qd} after accounting for the differing
    Debye mass of $\Nfour$ SYM.
    }
So the trivial configuration (\ref {eq:trivial})
describes a deconfined plasma phase
in which the center symmetry is spontaneously
broken while the $U(1)^3$ $R$-symmetry
(left invariant by the chemical potentials) is unbroken.

The free energy 
$
	F=-T\ln Z_{\text{ESYM}_3}
$,
where $Z_{\text{ESYM}_3}$ is the partition function of the
three-dimensional effective theory (\ref{LESYM}).
In the deconfined plasma phase
the order $\lambda^0$ and $\lambda^1$ contributions to $\beta F$
arise solely from the heavy modes which were integrated out
to produce the effective theory.
Hence, these contributions are completely contained in
the coefficient of the unit operator, $f$,
given in Eq.~(\ref{coeffidentity}).
The next contribution is of order $\lambda^{3/2}$ (not $\lambda^2$),
and comes from static fluctuations of the scalar fields
$\tilde \Phi_A$ and $\tilde A_0$.
In the full theory,
evaluation of the order $\lambda^{3/2}$ contribution requires an infinite
resummation of ring diagrams.
But this becomes an easy one-loop calculation using
the high temperature effective theory.

Because the spatial gauge fields $\tilde A_i$ are massless,
their one-loop contribution vanishes (using dimensional continuation).
The effective scalar fields $\tilde A_0$ and $\tilde \Phi_A$
each make a one-loop contribution to the free energy density of the form
\begin {equation}
    \half \Nc^2 \, T \int \frac{d^3p}{(2\pi)^3} \> \ln (p^2 + m^2)
    =
    -\Nc^2 \, T \> \frac{m^3}{12\pi} \,,
\label{eq:3dloopint}
\end {equation}
where the integral is evaluated using dimensional continuation,
and $m$ is the appropriate mass parameter for the field.
So the entire $\mathcal O(\lambda^{3/2})$ contribution to the
free energy density is
\begin{align}
	-\frac{\Nc^2 \, T}{12\pi}\Bigl(M_D^3+\sum_{A=1}^6 m_A^3\Bigr) \,.
\end{align}
Inserting the explicit expressions (\ref {scalarmass})--(\ref {eq:M_D})
for these masses, and adding ($T$ times) the coefficient $f$ of the unit
operator (\ref {coeffidentity}),
we obtain the free energy in the deconfined plasma phase,
through order $\lambda^{3/2}$,
\begin{subequations}\label{eq:SymmFreeEnergy}
\begin{align}
	F_{\rm plasma}=-\frac {\Nc^2}{12} \> T^4 \, \mathcal V \,
	\bigg\{2\pi^2
	&-3\lambda
	-\frac{3}{T^2R^2}
	+\sum_{p=1}^3\frac{2\mu_p^2}{T^2}
	+\sum_{i=1}^4\frac{\tilde{\mu}_i^2}{T^2}
\nonumber \\
	&+\frac {(2\lambda)^{3/2}}\pi
	+\frac 2\pi
	\sum_{p=1}^3
	    \left(\lambda+\frac{1}{T^2R^2}-\frac{\mu_p^2}{T^2}\right)^{3/2}
	+\mathcal{O}(\lambda^2)\bigg\} \;,
\end{align}
\end{subequations}
with $\mathcal V \equiv 2\pi^2R^3$ the spatial volume.
Note that this expression is only valid (and real)
when the chemical potentials satisfy the inequality (\ref{eq:mumax}).

\subsection{Is There a Higgs Phase?}\label{subsec:higgs}

Now consider the situation when the maximal chemical potential
exceeds $\lambda T^2 + R^{-2}$, so that some of the effective scalar
masses $m_A^2 = \lambda T^2 + R^{-2} - \mu_A^2$ become negative.
This makes the quadratic terms in the ESYM$_3$ scalar potential unstable,
so the minimum of the effective theory
scalar potential can no longer lie at the origin.
However, one may verify that
the sum of the tree (\ref{eq:Vtree}) and one-loop
(\ref{ESYMquartic}) quartic contributions to the scalar potential
is positive definite
for all non-zero $\tilde \Phi_A$,
showing that the quartic thermal corrections to the effective potential
lift the flat directions which are present in the tree-level potential.

This suggests that thermal corrections to the scalar interactions
in the effective theory may stabilize the theory when chemical
potentials are large, generating
a non-trivial minimum of the effective scalar potential.
Such a non-trivial minimum would correspond to a Higgs phase
with spontaneously broken $R$-symmetry.

Assume (for the moment) that terms in the scalar potential
involving higher than quartic powers of the field are unimportant.
One can explicitly minimize
the quadratic plus quartic contributions to the ESYM$_3$ scalar potential.
At the minimum one is balancing an unstable tree-level quadratic term
against a one-loop quartic term, so the rescaled eigenvalues (\ref{eq:rho})
of the scalar fields $\tilde\Phi_A$ are of order
\begin{equation}
    (\rho_A^m)^2 \sim \frac {|m_A^2|}{\lambda T^2} \,.
\label{eq:rhomin}
\end{equation}
The corresponding value of the (truncated) potential, at its minimum,
scales as
\begin {equation}
    -\Nc^2 \, T^3
    \left[ \max_A \Bigl(-\frac{m_A^2}{\lambda T^2}\Bigr) \right]^2 .
\label{eq:quarticmin}
\end {equation}
However, minimizing the truncated scalar potential and adding its value
to the coefficient of the identity operator
(\ref{coeffidentity})
does not produce a valid approximation for the 
free energy and, as we will see,
this putative Higgs phase simply does not exist.

There are two problems with the above
scenario suggesting the presence of a Higgs phase.
First, we neglected higher than quartic terms
in the ESYM$_3$ potential (\ref{eq:Vflat}).
This is only acceptable if $(\rho_A^m)^2 \ll 1$ which,
given the above estimate,
requires $|m_A^2| \ll \lambda T^2$.
The curvature and chemical potential contributions to $m_A^2$ are,
by previous assumption, individually of order $\lambda T^2$.
So neglecting terms higher than quartic in the potential requires that
$m_A^2$ be parametrically smaller than its individual pieces.
In other words,
one must restrict attention to a ``fine-tuned'' region in the phase diagram
sufficiently close to the surface where an effective scalar mass
first turns negative.

Second, and more importantly, the above scenario was based on a
classical analysis of the action (\ref {LESYM}) of the effective theory
--- so it neglected the effects of fluctuations in the static ESYM$_3$ fields.
To see if this matters, let
\begin{equation}
    \tilde\Phi_A = \langle \tilde\Phi_A\rangle + \delta\tilde\Phi_A \,,
\label{eq:dPhi}
\end{equation}
with the mean value $\langle \tilde\Phi_A\rangle$ lying along
a flat direction of the tree-level potential.
That implies that the mean values are simultaneously diagonal,
$\langle (\tilde\Phi_A)_{mn}\rangle = (\pi T/g_3) \, \rho_A^m \, \delta_{mn}$
(up to an irrelevant gauge transformation).
Inserting the decomposition (\ref{eq:dPhi}) into the tree-level potential
(\ref {eq:Vtree})
leads to non-zero mass corrections for the off-diagonal components
of the scalar field and $\tilde A_0$ fluctuations,%
\footnote
    {
      We have omitted off-diagonal terms
      contained in $\delta V_{\rm tree}(\tilde\Phi_A)$
      which mix $\delta\tilde\Phi_A$ with $\tilde A_0$,
      or mix $\delta\tilde\Phi_A$ with $\delta\tilde\Phi_B$ (for $A \ne B$).
      These terms may be canceled by
      using the three dimensional version of the gauge fixing term
      (\ref{eq:R_xi}), with gauge fixing parameter $\xi=1$.
    }
\begin{equation}
	\delta V_{\rm tree}(\tilde\Phi_A)
	=
	\sum_{m , n} \>
	M_{mn}^2 \,
	\Bigl\{
	\bigl| (\tilde A_0)_{mn} \bigr|^2
	+
	\sum_B
	\bigl| (\delta\tilde\Phi_B)_{mn}\bigr|^2
	\Bigr\}
	+
	\mathcal O(\delta\tilde\Phi^3)
	+
	\mathcal O(\delta\tilde\Phi \tilde A_0^2 )
	\,,
\label{eq:offdiag}
\end{equation}
with
\begin{equation}
    M_{mn}^2
    \equiv
    \pi^2 T^2 \, \sum_A (\rho_A^m - \rho_A^n)^2 \,.
\label{eq:Mmn}
\end{equation}
And inserting (\ref{eq:dPhi}) into the scalar kinetic term produces
mass terms for the off-diagonal components of the static gauge field,
$
    \tr (D_i \tilde\Phi_A)^2
    =
    \sum_{m , n} \>
    M_{mn}^2 \,
    \bigl| (\tilde A_i)_{mn} \bigr|^2
    + \cdots \,.
$

If the (rescaled) diagonal components $\{ \rho_A^m \}$ are $\mathcal O(1)$,
then the mass terms (\ref {eq:Mmn}) for off-diagonal fields induced
by the mean values of the diagonal components will be of order $T^2$,
or large compared to
the $\mathcal O(\lambda T^2)$ thermal contributions
due to the non-zero frequency modes,
as well as the curvature and chemical potential induced mass terms
[which we have assumed to also be $\mathcal O(\lambda T^2)$].
It will be sufficient to focus on this regime;
if a Higgs phase does exist, then the typical size of $\rho_A^m$ 
will need to be of this order.
%
%
Consequently, if the scalar fields $\tilde \Phi_A$
have non-trivial mean values lying along flat directions of the
tree-level potential, then off-diagonal components of
the static fields will act like heavy degrees of freedom
(just like the non-static Matsubara modes)
and need to be integrated out before considering the dynamics
of the remaining ``light'' diagonal modes.

Doing so is straightforward;
the basic ingredient is just the three dimensional loop integral
(\ref {eq:3dloopint}).
The result is a ``Higgs-branch'' three-dimensional effective theory 
which we will term ``HSYM$_3$'' and whose Lagrange density has the form
\begin {equation}\label{HSYM}
    {\cal L}_{{\rm HSYM}_3}
    =
    \,\tr\! \left[
	    \coeff{1}{2} (\tilde F_{ij})^2
	    + (D_i \tilde A_0)^2 + M_D^2 \, \tilde A_0^2
	    + (D_i\tilde \Phi_A)^2
	    \right]
	    + \overline V(\tilde \Phi_A) \,,
\end {equation}
where the fields are now diagonal,%
\footnote
    {
    For configurations in which the scalar field $\tilde\Phi_A$ has
    degenerate eigenvalues
    (which may naturally be grouped together into blocks of equal
    eigenvalues), it is the block off-diagonal components of the
    static fields which acquire mass,
    and the block-diagonal components which remain light.
    For these exceptional configurations, the residual gauge group is
    larger than $[U(1)]^{\Nc-1}$.
    }
with gauge group $[U(1)]^{\Nc-1}$.
The scalar potential $\overline V(\tilde \Phi_A)$ for the
remaining diagonal components is
\begin {align}
    \overline V(\tilde \Phi_A)
    &=
    f
    + \tr (m_A^2 \, \tilde\Phi_A^2)
    + \delta V_{\rm flat}(\tilde\Phi_A)
    + \delta V_{\rm off-diag}(\tilde\Phi_A) \,,
\end {align}
where $f$ is the additive constant (\ref{coeffidentity})
due to the non-zero frequency modes,
$\delta V_{\rm flat}(\tilde\Phi_A)$ is the one-loop potential
(\ref {eq:Vflat}) induced by non-zero frequency fluctuations,
and%
\footnote
    {
      This is just the result (\ref{eq:3dloopint}),
      adapted to the differing masses (\ref {eq:Mmn}),
      and multiplied by 8 to account for the
      $8N_c^2$ (real) bosonic degrees of freedom ---
      6 from the scalars and two from the transverse
      components of the gauge field.
      (The gauge-fixing ghosts cancel the
      $\tilde A_0$ and longitudinal contributions.)
    }
\begin{equation}
    \delta V_{\rm off-diag}(\tilde\Phi_A)
    =
    - \coeff 23 \pi^2 \> T^3 \, \tr\!
	\bigg[
	  \Big(
		\frac{g_3^2}{\pi^2T^2}
		\sum_A \tilde\Phi_A^\text{adj} \tilde\Phi_A^\text{adj}
	    \Big)^{3/2}
	\bigg] .
\end{equation}
Writing this explicitly in terms of the rescaled eigenvalues $\{\rho_A^m\}$
of $\tilde\Phi_A$, the potential is
\begin{align}
	\overline{V}(\rho_A^m)
	=
	\half \pi^2 \, T^3 \sum_{m,n}
	\bigg[\,
	h(\rho^{mn}) + \sum_{A} \delta\hat{m}_A^2 \, (\rho_A^m{-}\rho_A^n)^2
	\bigg]
	\,,
\label{eq:Vbar}
\end{align}
where we have defined an rms measure of eigenvalue differences,
\begin{equation}
    \rho^{mn}
    \equiv
    \Big[
	\sum_A \big(\rho_A^m - \rho_A^n\big)^2
    \Big]^{1/2} \,,
\end{equation}
introduced the rescaled mass shift
\begin{equation}
	\delta\hat{m}_A^2 \equiv \frac {m_A^2}{\lambda T^2} - 1
	= \frac {R^{-2}-\mu_A^2}{\lambda T^2} \;,
\end{equation}
and defined the one-dimensional function
\begin{equation}
    h(v) =
	-\coeff 13
	+ v^2
	-\coeff{4}{3} \, v^3
	+(\ln 2) \, v^4
	+\sum_{l=3}^\infty 
			8 \, (1-4^{-l+2}) \,
			\frac{(2l{-}5)!!}{(2l)!!} \,
			\zeta(2l{-}3) \,
			(-v^2)^l \,.
\label{eq:h(v)}
\end{equation}
The additive constant $-\frac 13$ comes from $f$.
We have dropped subleading $\mathcal O(\lambda T^3)$ contributions
to $f$, as these are parametrically smaller
than the $\mathcal O(T^3)$ terms retained in $\overline V$.

Inserting the defining series for the zeta function and interchanging
summations allows one to re-express $h(v)$ in an alternative form
which is more convenient for examining its global behavior, namely
\begin{equation}
    h(v) = -\coeff 13 + v^2 -\coeff 43 \, v^3 
    + \coeff 83 \sum_{j=1}^\infty \> (-1)^{j-1} 
	\left[(j^2{+}v^2)^{3/2} -j^3 - \coeff 32 v^2 \, j \right] \,.
\label{eq:hdef}
\end{equation}
The function $h(v)$ is graphed in Figure \ref{fig:hplt}.
For large arguments, $h(v)$ approaches zero exponentially rapidly,%
\footnote
    {
    To derive this asymptotic form, it is convenient to rewrite
    the infinite sum in (\ref{eq:hdef}) as a contour integral,
    leading to
    $
	h(z)
	=
	-\frac 13 + v^2
	+\frac 43 \lim_{\epsilon\to 0}
	\int_C \frac {dz}{2i} \> \csc(\pi z)
	[
	    (z^2 + v^2)^{3/2}
	    -(z^2 + \epsilon^2)^{3/2}
	    -\frac 32 v^2 (z^2 + \epsilon^2)^{1/2}
	]
    $,
    where the contour $C$ encircles the real axis counterclockwise
    and the branch cuts of the integrand run outward from the branch
    points to $\pm i\infty$.
    Deforming the contour so that it wraps around the branch cuts,
    producing an integral of the discontinuity across the cuts,
    leads to the stated asymptotic form.
    }
\begin{equation}
    h(v) \sim -\frac 4{\pi^2} \, (2v)^{3/2} \, e^{-\pi v} \,.
\end{equation}
The approach to zero as $v\to\infty$
may be understood as a consequence of supersymmetry.
In the potential (\ref {eq:Vbar}), the argument $\rho^{mn}$ of
the function $h$ is the ratio of
the mass scale $M_{mn}$ for the off-diagonal fields to ($\pi$ times)
the temperature.
A large ratio is equivalent to sending
$T \rightarrow 0$ while holding this mass scale fixed.
(Since we are assuming $1/(TR)^2 \sim \mu^2/T^2 \sim \lambda$, this also
implies sending $R \rightarrow \infty$ and $\mu \to 0$.) At zero temperature,
supersymmetry requires that all loop contributions to the
effective potential cancel.
So the cancellation of contributions to the one-loop potential
as eigenvalue differences become large follows from the
cancellation of one-loop contributions as $T \to 0$.

\begin{FIGURE}[t]
{
\centerline{\scalebox{0.7}{\includegraphics{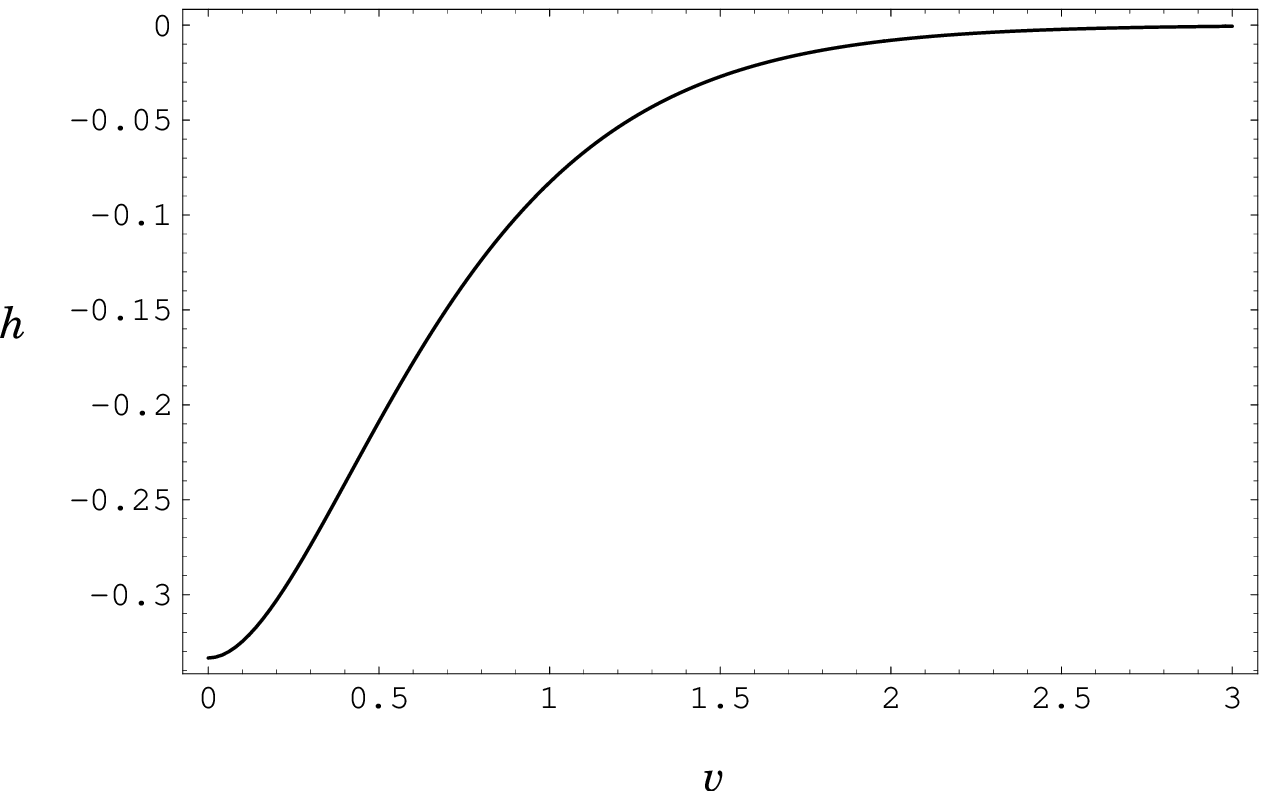}}}
\vspace*{-1cm}
\caption{\footnotesize 
  Graph of the function $h(v)$,
  defined in Eq.~(\ref {eq:hdef}).
}
\label{fig:hplt}
}
\end{FIGURE}

\begin{FIGURE}[t]
{
\centerline{\scalebox{0.9}{\includegraphics{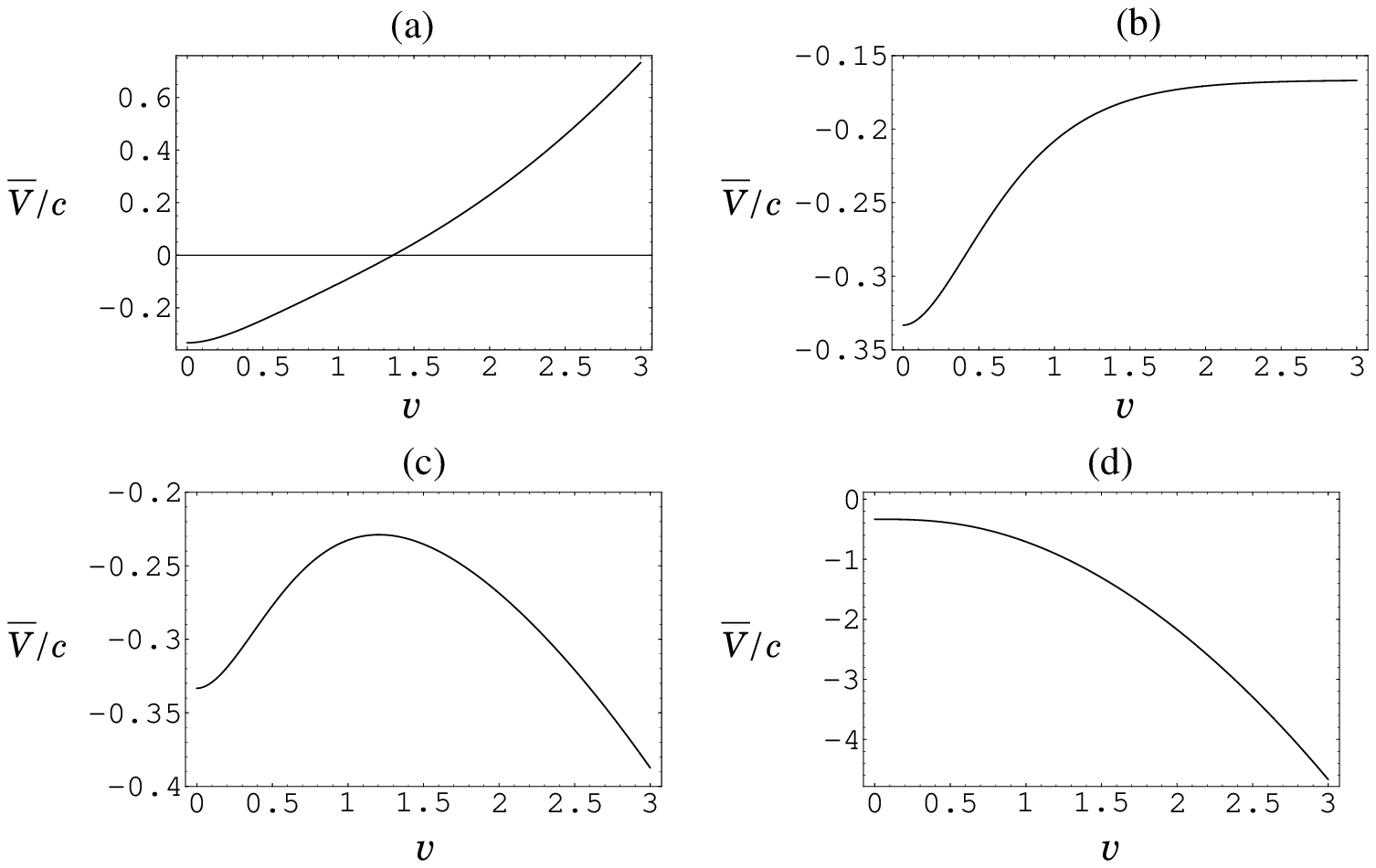}}}
\caption{\footnotesize 
  Plots of the scalar potential $\overline V$,
  divided by the overall factor
  $c \equiv \coeff 12 \pi^2 \Nc^2 T^3$,
  for the eigenvalue configuration
  $
      \rho_A^m= v \, \delta_A^1 \, (-1)^m
  $,
  as a function of $v$.
  The four plots correspond to
  $\delta\hat{m_A}^2=0.2$, $0.0$, $-0.05$ and $-1$, respectively.
  Note that half of the eigenvalues differences $\rho^{mn}$ are zero
  in this configuration,
  and each vanishing difference contributes
  $h(0) = -\frac 13$ to $\overline V/(\half \pi^2 T^3)$.
  This accounts for the non-zero asymptotic value in case (b).
  If $\delta\hat m_A^2$ is positive (or zero), then the potential is
  monotonically increasing with a unique minimum at the origin.
  If $-1 < \delta\hat m_A^2 < 0$, then the potential is
  unbounded below, with a local minimum at the origin.
  If $\delta\hat m_A^2 < -1$, then the potential is
  monotonically decreasing with a maximum at the origin.
}
\label{fig:potential}
}
\end{FIGURE}

\clearpage

As indicated in Eq.~(\ref{eq:Vbar}),
the complete potential $\overline V(\rho_A^m)$
is a sum of values of $h(v)$ plus
the sum of squares of the eigenvalue differences
weighted by the mass shift $\delta\hat m_A^2$.
Plots of the resulting potential, for a simple eigenvalue distribution
and several choices of mass shifts, are shown in Figure~\ref{fig:potential}.%
\footnote{
  Fig.~\ref{fig:potential} plots the potential when the eigenvalues
  of one scalar field are non-zero and all have the same
  magnitude, with half of them positive and half negative
  (so that the field is traceless).
}
Since the function $h(v)$ is bounded (for positive real $v$),
it is evident that the potential $\overline V(\rho_A^m)$
will be unbounded below if any mass shift $\delta\hat m_A^2$ is negative.
Since $\delta\hat m_A^2 \propto R^{-2} - \mu_A^2$, this means
that no truly stable equilibrium phase can exist if any chemical potential
exceeds $1/R$.
Moreover, the potential $\overline V(\rho_A^m)$ has no local minimum
for any non-zero set of eigenvalues $\{\rho_A^m\}$.
Hence there is no Higgs phase (not even a metastable Higgs phase) in this
weakly coupled, high temperature theory.

\subsection{High Density Metastability}\label{subsec:Instability}

When the largest chemical potential lies in the interval
\begin {equation}
    R^{-2} < \mu^2 < \lambda T^2 + R^{-2} \,,
\label{eq:metastable}
\end{equation}
the deconfined plasma phase
(corresponding to the origin of the field space) is locally stable,
but is not the global minimum of the free energy --- as there is no
global minimum.
For this range of chemical potential and finite $\Nc$, as we will see,
the deconfined plasma is
metastable with a lifetime which grows exponentially as $\Nc\to\infty$.
The onset of instability, at a (maximal) chemical potential of $1/R$,
is independent of $\lambda$ and thus exactly where the free theory
becomes ill-defined.
This may be understood as a consequence of the existence of
towers of BPS operators with vanishing anomalous dimensions
(independent of $\lambda$)
which map to towers of BPS states on $S^3$ with energies precisely equal
to their $R$-charge.
Consequently, the Boltzmann sum representation of the partition
function ceases to converge if any chemical potential exceeds $1/R$.
However, this does not prevent the existence of an arbitrarily long-lived
metastable phase above this threshold.

To estimate the decay rate when $\mu > 1/R$, consider the behavior of the
effective potential $\overline V(\tilde\Phi_A)$ for configurations
of the form
\begin{equation}
  \rho_A^m=
  v \, \delta_{A,1} \, (\delta_{m,1} - 1/N_c)
  \,,
\label{eq:rho-single}
\end{equation}
for some dimensionless real number $v$.
All but one of the eigenvalues remain near the origin in this configuration
as the single eigenvalue
$
    \rho_1^1 \equiv v \,
    (1{-}\frac 1\Nc)
$
is varied.
(This configuration satisfies the $SU(\Nc)$ tracelessness constraint,
$\sum_m \rho_A^m = 0$.)
Assume that $\mu_1$ is the largest chemical potential.
For this class of configurations, the potential $\overline V(\tilde\Phi_A)$
becomes
\begin{equation}
	\overline{V}(v)
	=
	-\coeff 16 \pi^2 T^3 \Nc^2
	+ \pi^2 T^3 \Nc \, [\, h(v)+\coeff 13 + \delta\hat m_1^2 \, v^2 ]
	+ \mathcal O(\Nc^0) \,.
\label{eq:delV-single}
\end{equation}
The interval (\ref {eq:metastable}) corresponds to
$\delta\hat m_1^2 \equiv (R^{-2} - \mu_1^2) / (\lambda T^2)$
lying between 0 and $-1$.
A plot of $\overline V(v)$ resembles case (c) of Fig.~\ref{fig:potential},
with a potential barrier separating the region near $v = 0$
from the bottomless region at large values of $v$.
However, for the eigenvalue configuration (\ref {eq:rho-single})
the height of this potential barrier
grows only linearly with $\Nc$,
\begin{equation}
    \Delta \overline V \equiv \max_v \overline V(v) - \overline V(0)
    = \mathcal O(\Nc \, T^3) \,.
\end{equation}
As seen from Eq.~(\ref {eq:Vbar}), this potential difference
receives contributions from every pair of
unequal eigenvalues of $\tilde\Phi_1$ and,
for this configuration, only $2(\Nc{-}1)$ terms involving the difference
between $\rho_1^1$ and the other $\Nc{-}1$ eigenvalues of $\tilde\Phi_1$
contribute.

The deconfined plasma phase corresponding to $v = 0$ can decay
via a thermal fluctuation in which a single eigenvalue
makes a large excursion from the origin to top of the barrier,
uniformly in space.
The probability for such a fluctuation has a Boltzmann suppression
factor of
\begin{equation}
    e^{-\mathcal V \, \Delta \overline V}
    =
    e^{-\mathcal O(\Nc \, (TR)^3)} \,,
\label{eq:Boltzmann1}
\end{equation}
where $\mathcal V  = 2\pi^2 R^3$ is the spatial volume.
(The usual factor of $\beta = 1/T$ appearing in
a Boltzmann factor was absorbed in our definition of
the scalar potential of the three dimensional effective theory.)

Alternatively, a thermal fluctuation could nucleate a 
critical bubble in which a single eigenvalue makes a large excursion
to values across the barrier with lower free energy density,
over a sufficiently large spatial
region so that bubble subsequently grows.
The process is characterized by a Euclidean bounce solution 
\cite{Coleman:1977py,Callan:1977pt}.
The required size of the critical bubble scales as $(\sqrt \lambda T)^{-1}$.
The action (in the effective theory) for such a solution is
$\mathcal O(\Nc/\lambda^{3/2})$, so the rate for this process will
have a Boltzmann suppression factor of%
\footnote
    {
      To see this, it is convenient to rescale spatial coordinates
      $\vec x \to \vec y / \sqrt\lambda T$, which
      corresponds to measuring distance in units of the Debye length.
      Then, for a single eigenvalue excursion, the scalar part
      of the action (\ref{HSYM}) reduces
      to a dimensionless action of the form
      $\int d^3y \> [\half (\nabla \rho)^2 + h(\rho)]$
      times an overall factor of $N_c/\lambda^{3/2}$.
      Therefore, the action of the bounce solution must scale in 
      the same way.
    }
\begin{equation}
    e^{-S_{\rm bounce}}
    =
    e^{-\mathcal O(\Nc \, \lambda^{-3/2})} \,.
\label{eq:Boltzmann2}
\end{equation}

When $1/(TR) = \mathcal O(\sqrt\lambda)$, which we have assumed,
then both processes will have rates which are exponentially
suppressed by an amount which scales as $\Nc / \lambda^{3/2}$.
Which process dominates depends on the pure numerical coefficients
in the exponents of (\ref{eq:Boltzmann1}) and (\ref{eq:Boltzmann2})
[which we have not evaluated] together with the value of $\sqrt\lambda \, TR$.%
\footnote
    {
    Note that decay rates
    via fluctuations involving large excursions
    of multiple eigenvalues are suppressed
    by additional exponentially small factors,
    relative to the rates of the single eigenvalue processes discussed above.
    As an extreme case, if all eigenvalues make equally large
    excursions, as in the configurations
    shown in Fig.~\ref{fig:potential}c, then
    the decay rate will be exponentially suppressed by a factor
    scaling as $N_c^2/\lambda^{3/2}$.
    }
Consequently, despite the fact that no true equilibrium state
exists when $\mu > 1/R$, as long as the maximal chemical potential
is within the interval (\ref {eq:metastable}), the deconfined plasma
phase will be metastable with a lifetime which diverges exponentially
as $\Nc \to \infty$.

The region of deconfined plasma metastability terminates when the
largest chemical potential reaches a maximal value,
\begin {equation}
    \mu_{\rm max}(T) = \sqrt { \lambda T^2 + R^{-2} }
    \times [1 + \mathcal O(\lambda)]
    \,,
\end {equation}
at which point the effective thermal mass of one or more of the
scalar fields becomes negative.
This is analogous to ``spinodal decomposition,'' which determines
the limit of supercooling or superheating at a typical first
order phase transition.
However, for weakly coupled $\Nfour$ SYM,
no new equilibrium state exists for $\mu > \mu_{\rm max}$.

\section{Summary and Comparison with Dual Gravitational Analysis}\label{comparison}

\begin{FIGURE}[ht]
{
\centerline{\scalebox{.75}{\includegraphics{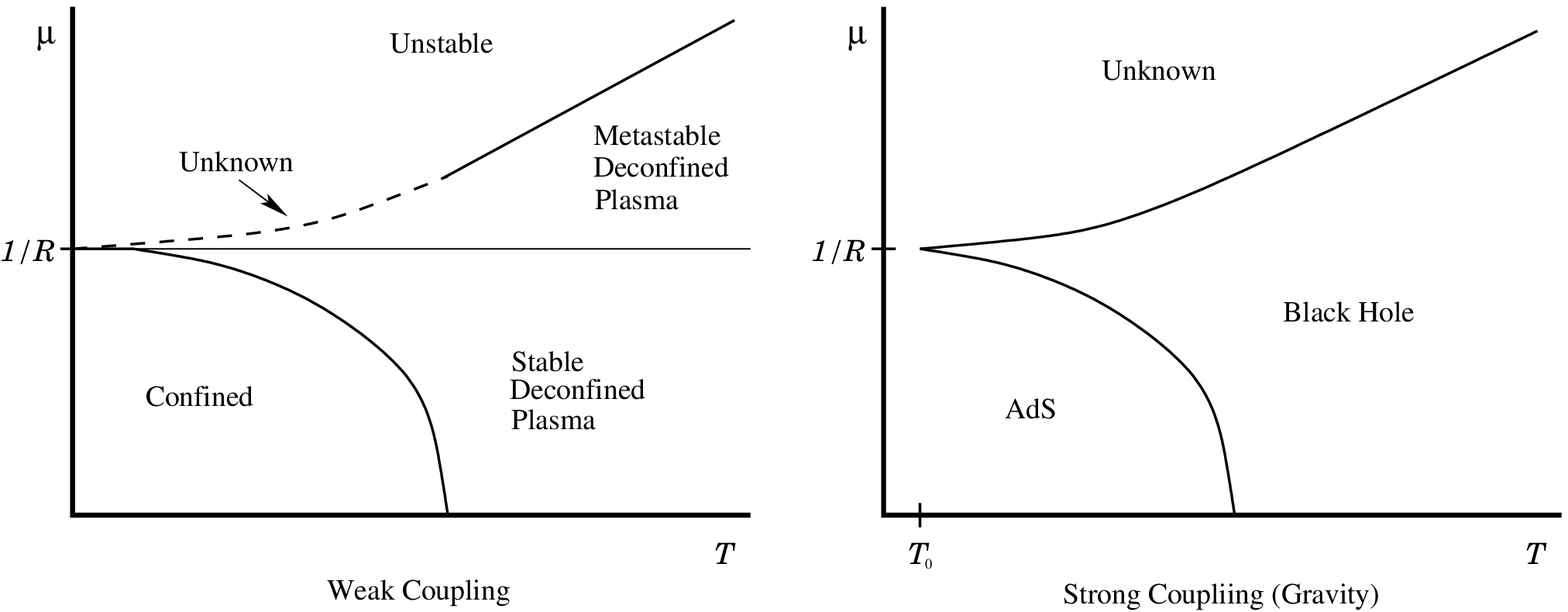}}}
\smallskip
\caption{\footnotesize 
  The phase structure of the weakly coupled gauge theory
  on the left, and of gravitational solutions to 5-$d$ supergravity,
  believed to correspond to the strongly coupled gauge theory,
  on the right.
  For both figures, $\mu$ is the largest of the three chemical potentials.
  In the weak coupling diagram,
  the location of the phase boundary forming the upper limit
  of the metastable plasma phase has not been evaluated at low temperature;
  this portion of the boundary is indicated by the dashed line.
  In the strong coupling diagram,
  $T_0$ is the temperature at which
  the black hole radius reaches to zero.
  Its value depends on the pattern of chemical potentials and equals
  $(\pi R)^{-1}$ for a single non-zero charge, while
  $T_0=0$ for three equal charges.
}
\label{fig:comparison}
}
\end{FIGURE}

The resulting phase diagram for weakly coupled $\Nfour$
supersymmetric Yang-Mills theory
is sketched on the left side of Figure~\ref{fig:comparison}.
A confinement/deconfinement transition line separates
a confining phase at lower temperature or chemical
potential from a deconfined plasma phase at higher
temperature or chemical potential.
The location of this line is known in the $\lambda \to 0$ limit
but, for the reasons discussed in section \ref{freetheory},
the order of this phase transition (at small but non-zero coupling)
is not currently known.
If the maximal chemical potential exceeds $1/R$,
then no truly stable equilibrium state exists.
But, at least for sufficiently high temperatures ($T \gg 1/R$), 
the deconfined plasma remains metastable beyond $\mu = 1/R$,
with a lifetime which grows exponentially as $\Nc \to \infty$.
This metastable region terminates at a spinodal decomposition
phase boundary when the largest chemical potential reaches
a limiting value $\mu_{\rm max}(T)$.
The $\mu = \mu_{\rm max}(T)$ boundary line asymptotically rises linearly with
temperature with slope $\sqrt\lambda$,
regardless of how many chemical potentials reach $\mu_c (T)$
simultaneously.

When $T \lesssim 1/R$,
the location of the spinodal decomposition phase boundary 
is not currently known (at non-zero but weak coupling).
Neither are order $\lambda$ corrections to the location
of the confinement/deconfinement transition line.
It is possible that the phase boundary remains separated from
the confinement/deconfinement transition line for all non-zero
temperatures, with both lines intersecting at $T=0$ and $\mu = 1/R$.
This possibility is sketched in Fig.~\ref{fig:comparison}.
But, for non-zero couplings,
the confinement/deconfinement transition line may instead
intersect the phase boundary line at a non-zero temperature.
In other words, the deconfined plasma region of the phase
diagram could pinch off before reaching zero temperature.
In this case, for sufficiently low temperature,
the confined phase might extend all the way to the
boundary of the phase diagram at $\mu = \mu_{\rm max}(T)$,
with no intervening phase transitions.
Which possibility occurs may well depend on
the particular pattern of chemical potentials
(for example, whether two or more chemical potentials coincide).

The right side of Fig.~\ref{fig:comparison} shows a sketch
of the behavior of gravitational solutions of five
dimensional $\mathcal{N}=2$ gauged supergravity,
which are believed to be related to the strong coupling limit
of $\Nfour$ supersymmetric Yang-Mills \cite{Chamblin:1999tk,Cvetic:1999ne}.
The similarities are obvious (in contrast to the
previous situation depicted in Fig.~\ref{IntroPic}).
A qualitatively similar confinement/deconfinement (or Hawking-Page)
transition line is present on both sides.
For the black hole, this transition is known to be first order.
The black hole instability line, where small perturbations
become thermodynamically unstable, appears completely analogous
to the spinodal decomposition phase boundary of the weak coupling
diagram.
The black hole instability line rises linearly
for temperatures large compared to the inverse of the
AdS$_5$ curvature radius, $1/R$,
but the slope depends on the pattern of chemical potentials
(unlike the case at weak-coupling).
With only one non-zero potential, for example, the asymptotic
slope is $\pi/\sqrt 2$, while for three equal chemical potentials
the slope is $2\pi$.

As noted in footnote \ref {fn:BHinstability} of the Introduction,
the temperature $T_0$ at which the Hawking-Page transition line
meets the black hole instability line 
depends on the pattern of chemical potentials.
For three equal chemical potentials (or equivalently three equal charges),
$T_0 = 0$.
With a single non-zero charge,
the temperature $T_0=(\pi R)^{-1}$.
In every case, the chemical potential at the intersection
of the instability and Hawking-Page transition lines is $1/R$.
This intersection corresponds to the point
where the black hole horizon radius shrinks to zero.
Since the phase diagram in the strong
coupling region is obtained by comparing 
the values of the action for black hole 
and AdS solutions at the same temperature,
the region of the phase diagram below $T_0$
is completely undetermined.

Ignoring the low temperature region
(where limited knowledge on both sides prevents a clear comparison),
there is one glaring difference between the weak and strong
coupling phase diagrams of Fig.~\ref{fig:comparison} ---
the non-perturbative metastability of the
deconfined plasma phase when the maximal chemical potential
exceeds $1/R$.
If there is a smooth interpolation between weak and strong coupling,
then there must be a similar non-perturbative black hole instability,
with an exponentially suppressed nucleation rate whose exponent
scales linearly with $\Nc$.
No such instability,
with an onset precisely at $\mu = 1/R$,
is currently known in the gravity dual.

\section{Outlook}\label{outlook}

There are a variety of possible extensions which should be feasible
and which would shed light on some of the issues discussed in this paper.

On the gravitational side, investigation into non-perturbative
instabilities of RN-AdS black holes (and their generalizations
to $\mathcal N=8$ gauged supergravity) is clearly needed.
Our weak coupling results,
plus presumed smooth interpolation between weak and strong coupling,
requires the existence of a non-pertur\-bative
black hole instability with an onset (at $\mu = 1/R$)
prior to the known point of perturbative instability.

Under gauge/string duality,
the eigenvalues of $\mathcal{N}=4$ SYM scalar fields are
identified with the locations of D3-branes in Type IIB string theory.
The $R$-charged black hole solutions arise from spinning D3-branes
\cite{Cvetic:1999ne}.
This suggests that the string theory manifestation
of the non-perturbative gauge theory instability should involve
a stack of D3-branes, for sufficiently high spin, splitting up 
into widely separated branes.
For $\mu > 1/R$, the force between branes at sufficiently large
separation should become repulsive.
Since the dominant instability in the gauge theory involves
a large excursion of a single eigenvalue, the dominant
high-spin D3-brane instability should involve a single brane
separating from the rest of the stack.
It should be possible to verify this scenario with a probe-brane
analysis in the RN-AdS background.

Alternatively, one may consider the analogue of non-perturbative
gauge theory instabilities involving large excursions by
multiple eigenvalues, namely
a process in which a stack of $\Nc$ coincident D3-branes
separates into a multi-stack configuration, with each stack containing
an order one fraction of the $\Nc$ branes. 
(Fig.~\ref{fig:potential}c and d correspond to the simplest case of
a symmetric two-stack instability.)
Though subdominant, this decay process should be
well described by supergravity approximations.

Finding such an instability for spinning D3-branes,
with the expected onset threshold of $\mu = 1/R$,
would provide further evidence supporting a smooth interpolation
between weak and strong coupling in $\Nfour$ supersymmetric Yang-Mills
theory (and the validity of AdS/CFT duality).
It might also clarify the puzzle, mentioned in Section~\ref{comparison},
involving the behavior of the gravitational system with large but unequal
chemical potentials, at temperatures below $T_0$ (where the horizon radius
of the $R$-charged black hole shrinks to zero).
At the moment, no upper limit on the chemical potential
of the gravitational phase diagram in this regime in known.

Turning now to the weak coupling analysis,
the order of the confinement/deconfine\-ment
phase transition at non-zero coupling remains uncertain,
for the reasons discussed in Section~\ref{freetheory}.
Its determination requires a three-loop computation similar to
(but more complicated than)
the one carried out for large $N_c$ pure Yang-Mills theory in 
Ref.~\cite{Aharony:2005bq}.
This computation, if carried out for non-zero chemical potentials,
would also allow one to determine how weak coupling
corrections shift the location of the transition line,
which is of particular interest at low temperature
where the transition line approaches the spinodal decomposition line.
If the interpolation between weak and strong coupling in $\Nfour$ SYM
is smooth, then the confinement/deconfinement transition is expected
to be first order.
Confirming (or refuting) this directly would clearly be desirable.

Finally, our weak coupling analysis focused on the
high temperature regime, $T \gg 1/R$.
Extending the treatment to $T \lesssim 1/R$ would be valuable.
For $T \ll 1/R$ one has a large time circle and a small spatial sphere,
so the non-zero angular momentum modes on the three-sphere
are heavy and may be integrated out, leading to an effective
one dimensional quantum mechanics.
Analysis of the resulting effective theory should reveal the
behavior of the spinodal decomposition phase boundary in the low
temperature region,
determine whether it intersects the confinement/deconfinement
transition line at a non-zero temperature $T_0$,
and perhaps even reveal a distinct high density,
low temperature phase.

\section*{Acknowledgments}

We would like to thank A.~Karch and S.~Minwalla for helpful discussions.
This work was supported, in part,
by the U.S. Department of Energy under Grant No.~DE-FG02-96-ER-40956.

\newpage

\appendix

\section{Derivation of Matrix Model with Chemical Potentials}
\label{MatrixModelDerivation}

To reduce the partition function of the free theory
(constrained by Gauss' law) to the matrix model,
as sketched in section \ref{PFunc},
one first separates the zero mode of $A_0$ by replacing
\begin{equation}
	A_\nu(x)\longrightarrow \tilde{A}_\nu(x) + \delta_{\nu,0} \, a / g \;,
\end{equation}
where $a$ is a constant traceless Hermitian $N_c\times N_c$ matrix,
and $\tilde{A}_0$ is orthogonal to the constant mode.
Without loss of
generality, we may assume that $a$ is diagonal and we denote its
entries (or eigenvalues) by $\{q^m\, |\, m=1,\cdots,N_c\}$ where the
$q^m$ are all real
and satisfy $\sum_mq^m=0$.
One plugs this shifted
form of the gauge field into the Lagrangian (\ref{fullLagrangian})
and then sends $g \to 0$.
It is convenient to choose
the gauge defined by the gauge-fixing term,
\begin{equation}\label{sec3gaugefixing}
	\tr\{(\partial_\nu \tilde A_\nu+i \, [a,\tilde A_0])^2\} \;.
\end{equation}
Define, for convenience, $q^{mn}\equiv q^m-q^n$ and
$\Box^{mn}\equiv (\partial_\nu+iq^{mn}\delta_{\nu,0})^2$.
The Lagrangian of the free theory may be rewritten as
the quadratic form
\begin{align}
	\hat{\mathcal{L}}_{\text{quad}}=
		\sum_{m,n}&\bigg\{(\tilde A_\nu)^*_{mn}(-\Box^{mn})(\tilde A_\nu)_{mn}
\nonumber\\
		&+((X_p)^*_{mn},(Y_p)^*_{mn})
			\begin{pmatrix}
			-\Box^{mn}+1-\mu_p^2 & 
			2i\mu_p(\partial_0+iq^{mn})	\\
			-2i\mu_p(\partial_0+iq^{mn})\; &
			-\Box^{mn}+1-\mu_p^2
			\end{pmatrix}
			\begin{pmatrix}
			(X_p)_{mn} \\
			(Y_p)_{mn}
			\end{pmatrix}
\nonumber\\
		&+(\bar{\psi}_i)^*_{mn}( i \Slash{\partial} - \gamma_0\,q^{mn}
		- i \tilde \mu_i \gamma_0\gamma_5)(\psi_i)_{mn}	\bigg\}
								\;,
\label{eq:Lhat}
\end{align}
where the color indices $m,n$ are now explicit,
with
$(\tilde A_\nu)_{mn}\equiv \tilde A_\nu^a \, (T^a)_{mn}$
and likewise for the other fields.

Table \ref{table:harmonics} in section \ref{PFunc} lists the 
eigenvalues of the spatial small fluctuation operators
whose eigenfunctions are (scalar, vector, or spinor)
spherical harmonics. Denote these eigenvalues by
$\Delta_g^2$, $\Delta_s^2$ and $\Delta_f^2$ for the transverse spatial gauge,
conformal scalar, and spinor
fields, respectively. Denote the Matsubara frequencies by
$\omega_k$,
for either bosons or fermions (as determined by context).
Then the contribution to $\ln Z$ from the gauge bosons and ghosts
(equal to $-\half$ times the logarithm of their functional determinant) is
\begin{equation}
	\ln Z_{\rm g}
	=
	-\half\biggl[\sum_k\sum_{m,n}\tr \ln\{(\omega_k+q^{mn})^2+\Delta_g^2\}
		-\tr \ln (\omega_k^2+\Delta_g^2)\biggr]
									\;.
\end{equation}
The trace is over the space of transverse vector fields on $S^3$.
The contribution from gauge-fixing ghosts cancels the contributions from
$\tilde A_0$ and longitudinal part of the spatial gauge field,
thereby effectively leaving two transverse gauge field degrees of freedom.
(This factor of two is
included in the degeneracy factor for transverse vector fields
listed in Table \ref{table:harmonics}.)
The second term compensates for the fact that there are really
only $\Nc{-}1$ independent diagonal components of the gauge field,
not $\Nc$, because we are dealing with the gauge group $SU(\Nc)$.

Block-diagonalizing the linear operator in (\ref {eq:Lhat}) for the scalars
(by working in the eigenspaces of $\partial_0$ and the spatial Laplacian)
and taking the determinant in each block gives the scalar contribution of
\begin{align}
	\ln Z_{\rm s}
	=
	-\half\sum_k\sum_p\bigg[\sum_{m,n}&\tr \ln\Bigl[\{(\omega_k+q^{mn}+i\mu_p)^2
	+\Delta_s^2\}\{(\omega_k+q^{mn}-i\mu_p)^2+\Delta_s^2\}\Bigr]
\nonumber\\
	-{}& \tr \ln\Bigl[\{(\omega_k+i\mu_p)^2+\Delta_s^2\}
	\{(\omega_k-i\mu_p)^2+\Delta_s^2\}\Bigr]\bigg]
									\;.
\end{align}
Similarly, the fermions give
\begin{align}
	\ln Z_{\rm f}
	=
	\half\sum_k
	\sum_i\bigg[\sum_{m,n}&\tr \ln\Bigl[\{(\omega_k+q^{mn}+i\tilde{\mu_i})^2
	+\Delta_f^2\}\{(\omega_k+q^{mn}-i\tilde{\mu_i})^2+\Delta_f^2\}\Bigr]
\nonumber \\
	-{}& \tr \ln\Bigl[\{(\omega_k+i\tilde{\mu}_i)^2+\Delta_f^2\}
	\{(\omega_k-i\tilde{\mu}_i)^2+\Delta_f^2\}\Bigr]\bigg]
									\;.
\end{align}
In integrating out the fermions, it is essential to recall that $\psi_i$
is a Majorana fermion. So $\psi_i$ and $\bar\psi_i$ are not independent
and the Grassmann integral gives the Pfaffian of
$(\Slash{\partial}+i\gamma_0\,q^{mn} -\tilde \mu_i \gamma_0\gamma_5)$,
not the determinant.

The Matsubara frequency sums can be carried out
(most easily by differentiating with respect to $\Delta^2$,
performing the sum, and then integrating back in $\Delta^2$) and yield
\begin{subequations}
\begin{align}\label{gauge_ghost}
	\ln Z_{\rm g}
	&=
	-\half\sum_{m,n}\tr \ln\Bigr[\{1-e^{-\beta (\Delta_g+iq^{mn})}\}
				\{1-e^{-\beta (\Delta_g-iq^{mn})}\}\Bigl]
			+\tr \ln (1-e^{-\beta\Delta_g})	\,,
\\[5pt]
\label{bosons}
	\ln Z_{\rm s}
	&=
	-\sum_p\bigg[\half\sum_{m,n}
		\tr\ln\Bigl[\{1-e^{-\beta (\Delta_s+\mu_p+iq^{mn})}\}
		\{1-e^{-\beta (\Delta_s+\mu_p-iq^{mn})}\}
\nonumber\\[-5pt]
		&\kern 3.2cm {}\times
		\{1-e^{-\beta (\Delta_s-\mu_p+iq^{mn})}\}
		\{1-e^{-\beta (\Delta_s-\mu_p-iq^{mn})}\}\Bigr]	\nonumber	\\
		&\kern 2.2cm {}
		-\tr \ln\Bigl[\{1-e^{-\beta(\Delta_s+\mu_p)}\}
		\{1-e^{-\beta(\Delta_s-\mu_p)}\}\Bigr]\bigg]	\;,
\\[5pt]
\label{fermions}
	\ln Z_{\rm f}
	&=
	\sum_i\bigg[\half\sum_{m,n}
	    \tr \ln\Bigl[\{1+e^{-\beta (\Delta_f+\tilde{\mu_i}+iq^{mn})}\}
		\{1+e^{-\beta (\Delta_f+\tilde{\mu_i}-iq^{mn})}\}
\nonumber\\
		&\kern 2.9cm {}\times
	    \{1+e^{-\beta (\Delta_f-\tilde{\mu_i}+iq^{mn})}\}
	    \{1+e^{-\beta (\Delta_f-\tilde{\mu_i}-iq^{mn})}\}\Bigr]
\nonumber\\
		&\kern 2cm {}
		-\tr \ln\Bigl[\{1+e^{-\beta(\Delta_f+\tilde{\mu}_i)}\}
		\{1+e^{-\beta(\Delta_f-\tilde{\mu}_i)}\}\Bigr]\bigg]	\;.
\end{align}
\end{subequations}
In these expressions, we have dropped temperature independent terms that
do not contribute to the thermodynamics of the theory.
To evaluate the remaining traces
(over transverse vector, scalar, or spinor fields on $S^3$),
it is convenient to expand each logarithm in a power series.
For example (taking part of the scalar contribution)
\begin{align}
	-\sum_p&\left[\half\sum_{m,n}\tr \ln\{1-e^{-\beta(\Delta_s+\mu_p+iq^{mn})}\}
			-\half\tr \ln\{1-e^{-\beta(\Delta_s+\mu_p)}\}\right]
\nonumber\\
		={}&\half\sum_p\sum_{h=0}^\infty(h+1)^2\sum_{l=1}^\infty\frac{1}{l}\>
							e^{-l\beta\Delta_s}
							e^{-l\beta\mu_p}
							\left(\sum_{m,n}e^{-il\beta q^m}
							e^{il\beta q^n}-1\right)
\nonumber\\
		={}&\half\sum_{l=1}^\infty\frac{1}{l}
		\left(\sum_{h=0}^\infty(h+1)^2e^{-l\beta\Delta_s}\right)
		\left(\sum_pe^{-l\beta\mu_p}\right)
		\left(\sum_me^{-il\beta q^m}\sum_ne^{il\beta q^n}-1\right)
\nonumber\\
	={}&\half\sum_{l=1}^\infty\frac{1}{l}\frac{x^l+x^{2l}}{(1-x^l)^3}
	\left(\sum_px^{l\mu_p}\right)\bigl(\tr U^l\,\tr U^{\dagger l}-1\bigr)	\;,
\end{align}
where $x\equiv e^{-\beta}=e^{-1/T}$ and $U\equiv\exp [i\beta a]$
is an $N_c\times N_c$ unitary matrix.
Similarly, one can work out the other logarithms.
One finds
\begin{align}
	\ln Z_g &=
	\sum_{l=1}^\infty\frac{1}{l}\>\frac{6x^{2l}-2x^{3l}}{(1-x^l)^3}\,
	\bigl(\tr U^l\,\tr U^{\dagger l}-1\bigr)	\;,
\\
	\ln Z_s &=
	\sum_{l=1}^\infty\frac{1}{l}\>\frac{x^l+x^{2l}}{(1-x^l)^3}
	\left[\sum_{p=1}^3\>(x^{l\mu_p}+x^{-l\mu_p})\right]
	\bigl(\tr U^l\,\tr U^{\dagger l}-1\bigr) \;,
\\
	\ln Z_f &=
	\sum_{l=1}^\infty\frac{(-1)^{l+1}}{l}\>\frac{2x^{\frac{3}{2}l}}{(1-x^l)^3}
	\left[\sum_{i=1}^4\>(x^{l\tilde{\mu}_i}+x^{-l\tilde{\mu}_i})\right]
	\bigl(\tr U^l\,\tr U^{\dagger l}-1\bigr)
\nonumber\\
	&=\sum_{l=1}^\infty\frac{(-1)^{l+1}}{l}\>\frac{2x^{\frac{3}{2}l}}{(1-x^l)^3}
	\left[\prod_{p=1}^3\>(x^{\frac{1}{2}l\mu_p}+x^{-\frac{1}{2}l\mu_p})\right]
	\bigl(\tr U^l\,\tr U^{\dagger l}-1\bigr)	\;,
\end{align}
where the definition (\ref{mu_tilde}) of the effective fermion
chemical potentials $\tilde\mu_i$ was used in the last line.

Combining these terms and defining
\begin{equation}
    S_{\rm eff}(U)
    \equiv
    -\ln Z_g -\ln Z_s -\ln Z_f \,,
\end{equation}
gives the result quoted earlier in (\ref {eq:effS})
and (\ref{singlepp}).
The exponential $e^{-S_{\rm eff}(U)}$
gives the result of integrating out all fields
except the zero mode of $A_0$.
What remains is a single integral over the matrix $U$,
as shown in (\ref{matrixZ}).

\section{One-Loop Matching for ESYM$_3$}\label{Diagram}

\subsection{Coefficient of the Identity Operator}\label{IDOp}

The leading contribution to the coefficient of the identity
operator in ESYM$_3$ may be obtained by computing
the functional determinants which result from integrating out the
heavy modes, neglecting interactions.
To evaluate the contributions 
from the heavy bosonic modes, first consider an
$O(2)$ invariant theory (in flat space) with two real massive scalar fields.
One may easily evaluate the contribution to the free energy produced
by integrating out these fields.
With a non-zero chemical potential, the high temperature expansion
of the result is \cite{Haber:1981ts}
\begin{equation}
	\frac{F}{\cal V}=-\frac{\pi^2}{45}\,T^4+\frac{m^2{-}2\mu^2}{12}\,T^2
			-\frac{(m^2{-}\mu^2)^{3/2}}{6\pi}\,T+\cdots 
								\;,
\end{equation}
where $\cal V$ is the volume of the space.
This result contains the contributions of both the heavy non-zero
Matsubara modes and the light zero-frequency mode.
For our purposes, the light mode 
contribution must be subtracted out.
The zero Matsubara mode contribution
to the above expression is given by the integral,
\begin{equation}
	\frac{F^{n=0}}{\cal V}
	=
	T\int\frac{d^3p}{(2\pi)^3}\> \ln ({p}^2+m^2-\mu^2)
	=
	-\frac{(m^2{-}\mu^2)^{3/2}}{6\pi} \, T	\;.
\end{equation}
This is exactly the third term in $F/\cal V$.
This was inevitable ---
such a non-analytic term is related to the infrared behavior of the theory
and can only come from fluctuations on scales large compared to $1/T$.
Therefore, the heavy mode
contribution is
\begin{equation}
	\frac{F^{n\neq 0}}{\cal V}
		=-\,T^4\left\{\frac{\pi^2}{45}-\frac{1}{12}\left(\frac{m^2
							{-}2\mu^2}{T^2}\right)
	+\mathcal{O}\left(\frac{\mu^2m^2}{T^4},\frac{m^4}{T^4},\frac{\mu^4}{T^4}
	\right)\right\} .
\end{equation}
In $\Nfour$ SYM, we have $3\Nc^2$ such pairs of scalars. Moreover, gauge bosons
and ghosts contribute $\Nc^2$ times the above expression with $m=\mu=0$.
Hence the total bosonic contribution is,
\begin{equation}
	\frac{F^{n\neq 0}_{B}}{\cal V}
		=-\,\Nc^2T^4\left\{\frac{4\pi^2}{45}-\frac{1}{12}\left[
				\frac{3}{T^2R^2}
				-\frac{2(\mu_1^2+\mu_2^2+\mu_3^2)}{T^2}\right]
	+\mathcal{O}(\lambda^2)\right\} ,
\end{equation}
where we have set $m^2=1/R^2$. Because we are working in the regime where
$
	1/(TR)^2\sim\mu^2/T^2\sim\lambda
$,
this result is correct through $\mathcal O(\lambda^2)$.

The analogous contribution from
one Majorana fermion with chemical potential $\mu$
[introduced as in Eq.~(\ref {eq:weyl to majorana})]
is
\begin{equation}
	\frac{F}{\cal V}=-\, T^4\left(\frac{7\pi^2}{360}
			+\frac{\mu^2}{12\, T^2}
			+\frac{\mu^4}{24\, T^4}\right) .
\end{equation}
Since fermions do not have zero-frequency modes, there is no light mode
contribution to subtract.
We have altogether $4\Nc^2$ Majorana fermions with four
different chemical potentials,
so the total one-loop fermion contribution is
\begin{equation}
	\frac{F_{F}}{\cal V}=-\Nc^2T^4\left\{\frac{7\pi^2}{90}
	    +\frac{1}{12}\left(\frac{\tilde{\mu}_1^2+\tilde{\mu}_2^2
			    +\tilde{\mu}_3^2+\tilde{\mu}_4^2}{T^2}\right)
	    +\mathcal{O}(\lambda^2)\right\} .
\end{equation}

Combining the boson and fermion results,
we obtain the one-loop contribution
to the free energy density,
expanded in powers of $1/(TR)$ and $\mu/T$ through $\mathcal{O}(1/T^2)$,
\begin{equation}
	\frac{F_{\rm 1-loop}}{\cal V}
	=
	-\frac{\Nc^2T^4}{12}\left\{2\pi^2
		-\frac{3}{T^2R^2}
		+\frac{2(\mu_1^2+\mu_2^2+\mu_3^2)}{T^2}
		+\frac{\tilde{\mu}_1^2+\tilde{\mu}_2^2
				+\tilde{\mu}_3^2+\tilde{\mu}_4^2}{T^2}
		+\mathcal{O}(\lambda^2)\right\} .
\end{equation}
However, this is not a complete result.
Because we are interested in the regime where
$1/(TR)^2\sim\mu^2/T^2\sim\lambda$,
two-loop contributions will yield $\mathcal O(\lambda)$
corrections to the free energy density which are comparable
in size to the curvature and chemical potential dependent parts
of the one-loop result.
The required two-loop contribution may be evaluated
ignoring curvature and chemical potential corrections altogether,
since these will only give subleading
$ \mathcal O(\lambda / (TR)^2)$ or $\mathcal O(\lambda \mu^2/T^2)$
terms, which are both order $\lambda^2$.
One finds \cite{Kim:1999sg,Fotopoulos:1998es,Vazquez-Mozo:1999ic},
\begin{equation}\label{twoloop}
	\frac{F_{\rm 2-loop}}{\cal V}
	=
	\frac{1}{4}\, \Nc^2T^4\, \lambda	
	    +\mathcal{O}(\lambda^2) \,.
\end{equation}
With massless fields in flat space, there are no light-mode contributions
which need to be subtracted out.
Thus, the coefficient of the identity operator in ESYM$_3$,
equal to $\beta$ times the free energy density due to the
heavy modes, is given by
\begin{align}
	f=-\frac{\Nc^2T^3}{12}\bigg\{2\pi^2
	-3\lambda
	-\frac{3}{T^2R^2}
	&+\frac{2(\mu_1^2+\mu_2^2+\mu_3^2)}{T^2}
	+\frac{\tilde{\mu}_1^2+\tilde{\mu}_2^2
				+\tilde{\mu}_3^2+\tilde{\mu}_4^2}{T^2}
				+\mathcal{O}(\lambda^2)\bigg\} \,.
\end{align}

\subsection{Scalar Thermal Mass}\label{ScalarMass}

\begin{FIGURE}
{
\centerline{\scalebox{.65}{\includegraphics{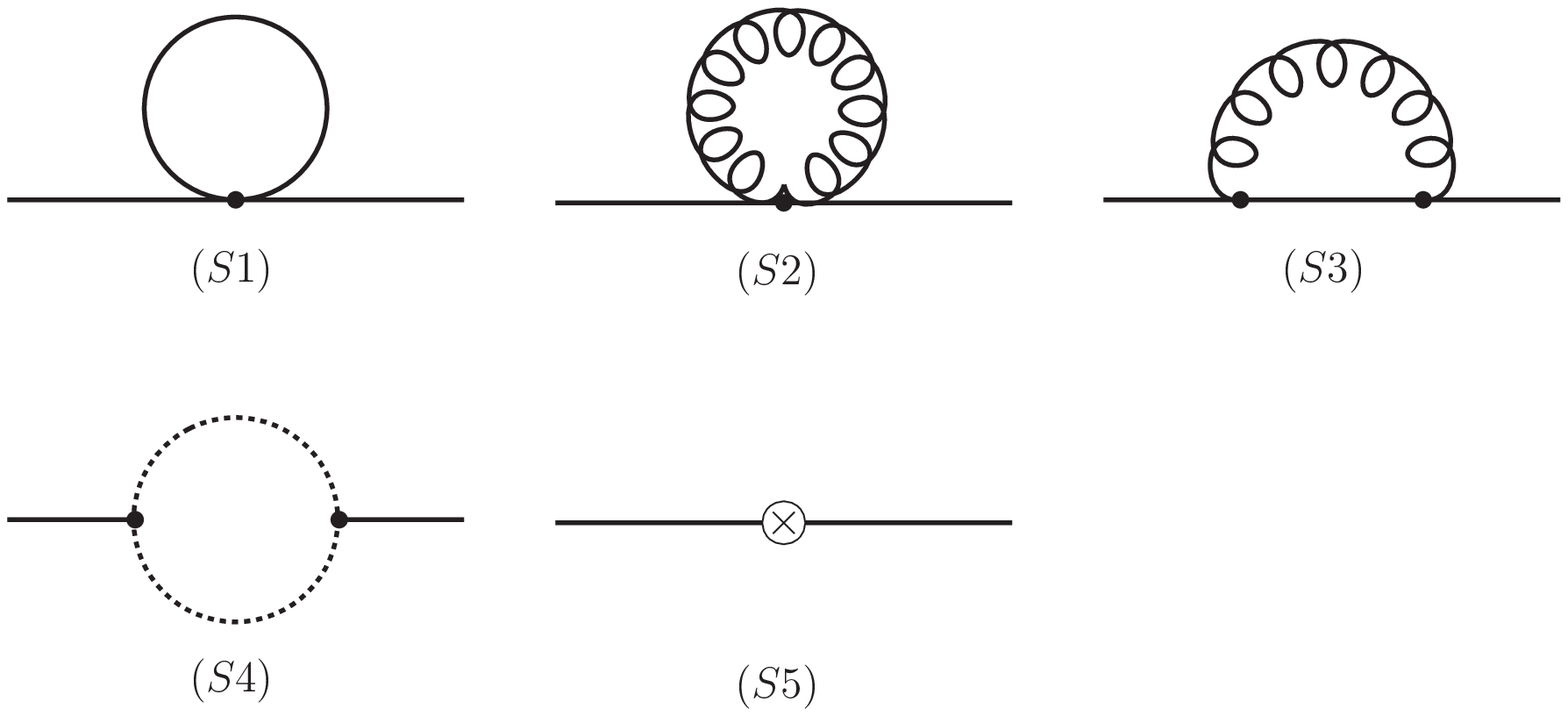}}}
\vspace*{-1em}
\caption{\footnotesize One-loop self-energy diagrams for the scalar fields.
The solid lines are scalars,
curly lines are gauge bosons, and dotted lines are Majorana fermions.
The final diagram $(S5)$ 
denotes the wavefunction renormalization counterterm. The external lines are
light modes with small spatial momenta, while the fields running inside
the loops are heavy modes.}
\label{selfenergy}
}
\end{FIGURE}

The lowest-order thermal mass correction for the scalar fields
comes from the self-energy diagrams shown in Fig.~\ref{selfenergy}.
As noted in section \ref {hightemp},
to extract the $\mathcal O(\lambda T^2)$ contribution to the mass (squared),
one may evaluate these diagrams in flat space and without
chemical potentials.
Including curvature and chemical potential corrections in these diagrams
would give subleading $\mathcal O(\lambda^2)$ terms,
which are beyond the accuracy we need.
Since we are interested in the construction of ESYM$_3$,
only the heavy mode contributions to each loop should be included.
To obtain the thermal mass, it is sufficient to evaluate
these diagrams at zero external momentum.
But we will keep the external (Euclidean) momentum $p = (0,\vec p\,)$
non-zero (but small compared to $T$) in order to extract the
one-loop wavefunction renormalization for the scalars,
which will be useful in the next subsection.
We use dimensional continuation and the $\overline{MS}$
renormalization scheme,
and denote the arbitrary renormalization scale by $\Lambda$.
Integrals are carried out in the standard manner with the convention that
\begin{equation}\label{defsumint}
	\sumintp \equiv  \left(\frac{e^{\gamma_E}\Lambda^2}{4\pi}\right)^{\epsilon}T\sum_{n\neq 
			0}\int\frac{d^{3-2\epsilon}p}{(2\pi)^{3-2\epsilon}} \;.
\end{equation}
The prime in the left-hand side
signifies the omission of the zero frequency mode
in bosonic loop integrals.
We use a Lorentz gauge-fixing term with arbitrary gauge parameter $\alpha$.
The dependence on $\alpha$ will drop out of the result for the thermal mass
correction, but keeping $\alpha$ arbitrary serves as a useful check on the
calculation.

Evaluating the diagrams of Fig.~\ref{selfenergy} is straightforward
and one finds:
\begin{subequations}
\begin{align}
	(S1)&=-\frac{5}{12}\, \lambda T^2	\,,\\
	(S2)&=-\frac 1{12}\, (3+\alpha)\, \lambda T^2	\,,\\
	(S3)&= \frac{\alpha}{12}\, \lambda T^2
		+ \frac{\lambda}{16\pi^2} \,
		    \Bigl( \frac{1}{\epsilon}+L_b\Bigr) \,
		    (3{-}\alpha) \, p^2
		+ \mathcal O\Bigl(\frac{p^4}{T^2}\Bigr) \,,\\
	(S4)&= -\frac{1}{3}\, \lambda T^2
		-\frac{\lambda}{4\pi^2}
		    \Bigl(\frac{1}{\epsilon}+L_f\Bigr) \, p^2
		+ \mathcal O\Bigl(\frac{p^4}{T^2}\Bigr) \,,\\[5pt]
	(S5)&=-\delta Z^{(1)} \, p^2		\,,
\end{align}
\end{subequations}
where
$\delta Z^{(1)}$ is the one-loop wavefunction renormalization counterterm.
We have defined
\begin{equation}\label{LbLf}
	L_b\equiv \ln\frac{\Lambda^2}{T^2}-2\ln (4\pi)+2\gamma_E	
								\;,\quad
	L_f\equiv \ln\frac{\Lambda^2}{T^2}-2\ln (4\pi)+2\gamma_E +4\ln 2	
								\;,
\end{equation}
with $\gamma_E$ is Euler's constant. 
The sum of these diagrams gives (minus) the hard mode contribution to
the one-loop self-energy,
\begin{align}
	-\Pi_{\rm hard}(p) &\equiv  (S1)+(S2)+(S3)+(S4)+(S5)
\nonumber\\
	&=-\lambda T^2
	+p^2\frac{\lambda}{16\pi^2}\,
	    \left\{ (3{-}\alpha)L_b-4L_f -\frac {1{+}\alpha}{\epsilon}
	    \right\}
	-p^2 \, \delta Z^{(1)}
	+ \mathcal O\Bigl(\frac{p^4}{T^2}\Bigr) \,.
\end{align}
Choosing
\begin{equation}\label{Z}
	\delta Z^{(1)}=-(1+\alpha) \, \frac{\lambda}{16\pi^2\epsilon}
\end{equation}
absorbs the logarithmic divergence.
As always, temperature independent renormalization suffices to
remove UV divergences at any temperature.
The hard-mode self-energy is thus
\begin{equation}
	-\Pi_{\rm hard}(p)=-\lambda T^2+p^2\frac{\lambda}{16\pi^2}
	\left\{(3{-}\alpha)L_b-4L_f\right\}
	+ \mathcal O\Bigl(\frac{p^4}{T^2}\Bigr) \,.
\end{equation}

The resulting renormalized scalar propagator,
in flat space and without chemical potentials,
at separations large compared to $1/T$, is
\begin{equation}\label{eq:GAB}
    \langle \Phi_A(x) \Phi_B(y) \rangle
    =
    \delta_{AB} \> T \int \frac {d^3p}{(2\pi)^3} \>
    e^{i \vec p \cdot (\vec x-\vec y)} \> G(p) \,,
\end{equation}
with
\begin{align}\label{eq:Ginv}
    G^{-1}(p)
    &=
    p^2 + \Pi_{\rm hard}(p) + \Pi_{\rm soft}(p)
\nonumber\\
    &=
    p^2 (1 + A) + \lambda\, T^2 + \Pi_{\rm soft}(p)
	    + \mathcal O(p^4/T^2) + \mathcal O(\lambda^2 T^2) \,,
\end{align}
where
$
    A \equiv 
    -  \frac{\lambda}{16\pi^2} \left\{(3{-}\alpha)L_b-4L_f\right\}
$
and $\Pi_{\rm soft}(p)$ is the self-energy contribution due to
soft modes.

In view of the relation (\ref {eq:rescale}),
the long distance correlator (\ref {eq:GAB}) should be matched with
$Z_3 T \langle \tilde \Phi_A(x) \tilde \Phi_B(y)\rangle$,
computed in the effective theory.
Again neglecting curvature and chemical potential corrections,
the effective theory correlator is
\begin {equation}
    \langle \tilde \Phi_A(x) \tilde \Phi_B(y)\rangle
    =
    \delta_{AB} \,
    \int \frac {d^3p}{(2\pi)^3} \>
    e^{i \vec p \cdot (\vec x-\vec y)} \> \widetilde G(p) \,,
\end {equation}
with
\begin{equation}
    \widetilde G^{-1}(p)
    =
    p^2 + \delta m^2(T) + \Pi_{\rm soft}(p) \,,
\end{equation}
and $\Pi_{\rm soft}(p)$ is the same soft mode contribution as in (\ref {eq:Ginv}).
Therefore, the parameters of the effective theory must be adjusted to make
$\widetilde G^{-1}(p) / Z_3$ coincide with $G^{-1}(p)$.
Through one-loop order, the required matching is
\begin{equation}\label{eq:Z3}
    Z_3 = (1 + A)^{-1}
	= 1 +  \frac{\lambda}{16\pi^2} \left\{(3{-}\alpha)L_b-4L_f\right\}
	+ \mathcal O(\lambda^2) \,,
\end{equation}
and
\begin{equation}
    \delta m^2(T) = \lambda \, T^2 + \mathcal O(\lambda^2 T^2) \,.
\end{equation}

\subsection{Scalar Quartic Coupling}\label{QuarticSubsection}

Calculating the one-loop thermal corrections to the scalar quartic interactions
which will appear in the high temperature effective theory
requires evaluation of the one-loop four-point correlator of light-mode 
scalar fields, at zero external momentum.
The analysis is similar to the computation by
Nadkarni \cite{Nadkarni:1988fh} of the quartic interactions of $A_0$
in EQCD.
The required diagrams are shown in Fig.~\ref{quartic}.
Once again, only heavy modes should
be regarded as running around the loops,
and one may work directly in flat space, ignoring
curvature and chemical potential corrections.

\begin{FIGURE}
{
\vspace{2em}
\centerline{\scalebox{.75}{\includegraphics{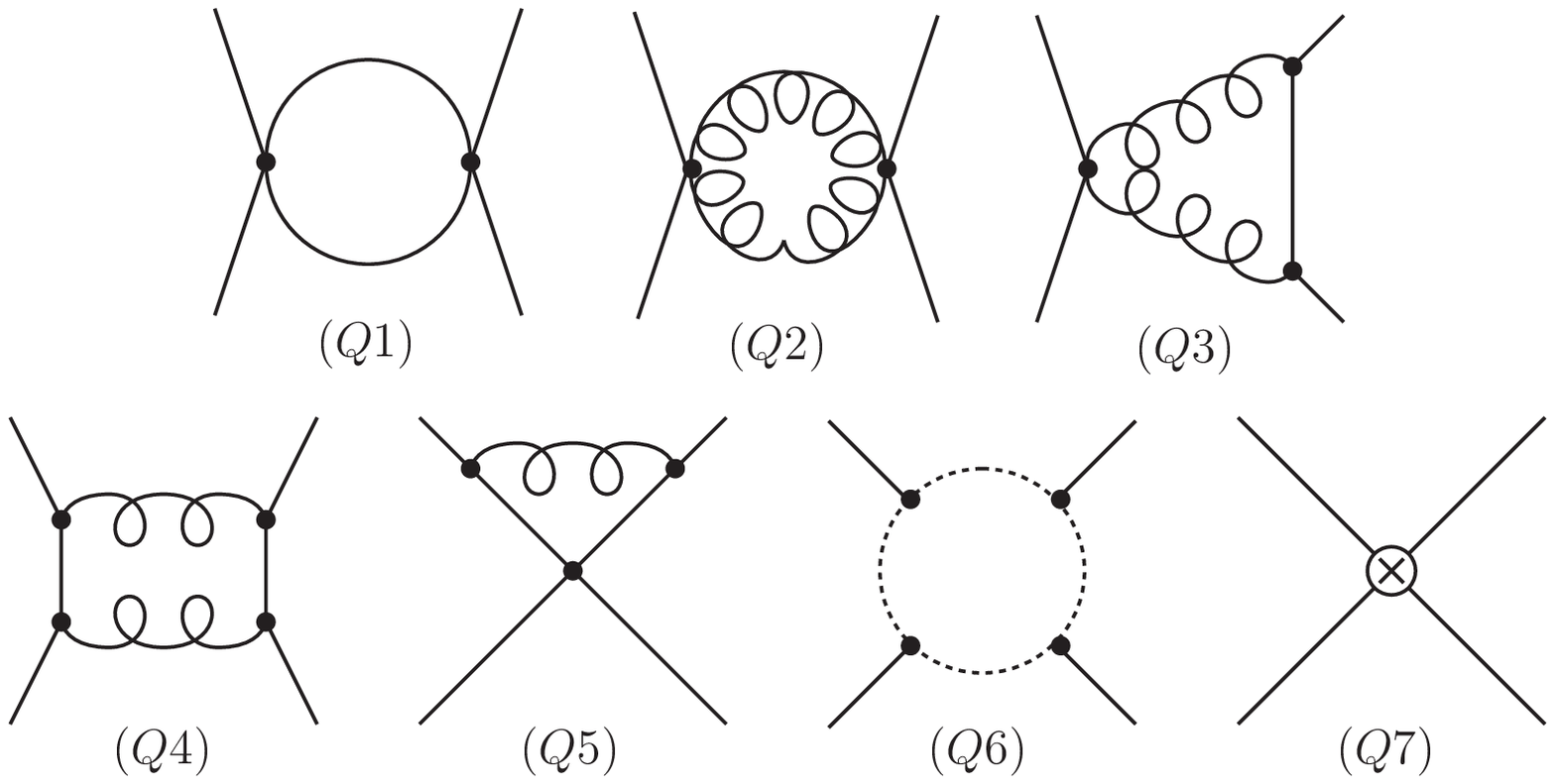}}}
\caption{\footnotesize
One loop diagrams which lead to quartic scalar interactions
in the high temperature effective theory.
As in Fig.~\ref{selfenergy}, solid, curly, and dotted lines
denote scalar, gauge boson, and fermion lines, respectively.
In Diagram $(Q6)$, it is most convenient to treat the external fields
$X$ and $Y$ separately due to their different Yukawa couplings.
Diagram $(Q7)$ denotes the contribution induced by wavefunction
renormalization.
Not shown explicitly are the various permutations of each diagram.
\label{quartic}
}}
\end{FIGURE}

We evaluate these diagrams using dimensional continuation and
$\overline{MS}$ renormalization, as described in appendix \ref {ScalarMass},
and an arbitrary gauge fixing parameter $\alpha$,
even though choosing Landau gauge ($\alpha\,{=}\,0$)
would reduce the number of the diagrams to be computed 
(namely diagrams $(Q3)$, $(Q4)$, and $(Q5)$ would all vanish at zero
external momentum).
Nevertheless, retaining an arbitrary gauge fixing parameter
in order to verify the cancellation of $\alpha$ dependence
in the resulting effective quartic coupling serves as a useful
consistency check.

To express the results compactly, we will use $\{V^a\}$ to denote
Hermitian $SU(\Nc)$ adjoint representation basis matrices, with%
\footnotemark
\begin{equation}\label{defT}
	(V^a)_{bc}\equiv -if^{abc} \;,
\end{equation}
and define the following index structures:
\begin{subequations}
\begin{align}
	(\Gamma_1)^{abcd}_{ABCD}\equiv N_c^{-2}
			\Bigl[\>
			\tr (V^aV^bV^cV^d)&\,
						(\delta_{AB}\,\delta_{CD}
						+\delta_{AD}\,\delta_{BC}
						-2\delta_{AC}\,\delta_{BD})	\nonumber\\
			+\tr (V^aV^bV^dV^c)&\,(\delta_{AC}\,\delta_{BD}
						+\delta_{AB}\,\delta_{CD}
						-2\delta_{AD}\,\delta_{BC})	\nonumber\\
			+\tr (V^aV^cV^bV^d)&\,(\delta_{AC}\,\delta_{BD}
						+\delta_{AD}\,\delta_{BC}
						-2\delta_{AB}\,\delta_{CD})
			\Bigr]\;,
\\[5pt]
	(\Gamma_2)^{abcd}_{ABCD}\equiv N_c^{-2}
			\Bigl[\>
			\tr (V^aV^bV^cV^d)&\,
						(\delta_{AB}\,\delta_{CD}
						+\delta_{AD}\,\delta_{BC})	\nonumber\\
			+\tr (V^aV^bV^dV^c)&\,(\delta_{AC}\,\delta_{BD}
						+\delta_{AB}\,\delta_{CD})	\nonumber\\
			+\tr (V^aV^cV^bV^d)&\,(\delta_{AC}\,\delta_{BD}
						+\delta_{AD}\,\delta_{BC})
			\Bigr]\;.
\end{align}
\end{subequations}
The tree-level quartic scalar vertex is
$
    -2\, \lambda \, (\Gamma_1)^{abcd}_{ABCD}\,
$.
\footnotetext
    {
    Useful adjoint representation trace identities include
    \begin{subequations}
    \begin {align}
	\tr(V^a V^b) &= \Nc \> \delta^{ab} \,,
    \\[7pt]
	\tr(V^a V^b V^c) &= -\coeff i2 \, \Nc \, f^{abc} \,,
    \\[7pt]\label{UsefulIdentity}
	\tr(V^a V^b V^c V^d) &= \Nc \, \tr(T^a T^b T^c T^d + T^b T^a T^d T^c)
	+ \half \, ( \delta^{ab} \delta^{cd} +
		     \delta^{ac} \delta^{bd} +
		     \delta^{ad} \delta^{bc} ) \,,
    \\[7pt]
	\tr(V^a V^b V^c V^d) &- \tr(V^b V^a V^c V^d)
	= -2\Nc \, f^{abe}f^{cde}  \,,
    \end{align}
    \end{subequations}
    where $\{T^a\}$ are the fundamental representation basis matrices.
    \vspace{2em}
    }

The diagrams of Figure \ref {quartic} can be computed in a
straightforward manner, and one finds
\begin{subequations}
\begin{align}
	(Q1)&=\frac{\lambda^2}{16\pi^2}\,
		    \Bigl(\frac{1}{\epsilon}+L_b\Bigr)
	    \left[4(\Gamma_1)^{abcd}_{ABCD}+5(\Gamma_2)^{abcd}_{ABCD}\right]
	    ,\\[5pt]
	(Q2)&=(3+\alpha^2)\,\frac{\lambda^2}{16\pi^2}\,
		    \Bigl(\frac{1}{\epsilon}+L_b\Bigr)\,
			(\Gamma_2)^{abcd}_{ABCD}	\;,\\[5pt]
	(Q3)&=-2\alpha^2\,\frac{\lambda^2}{16\pi^2}\,
		    \Bigl(\frac{1}{\epsilon}+L_b\Bigr)\,
			(\Gamma_2)^{abcd}_{ABCD}	\;,\\[5pt]
	(Q4)&=\alpha^2\,\frac{\lambda^2}{16\pi^2}\,
		    \Bigl(\frac{1}{\epsilon}+L_b\Bigr)\,
			(\Gamma_2)^{abcd}_{ABCD}	\;,\\[5pt]
	(Q5)&=-4\alpha\, \frac{\lambda^2}{16\pi^2}\,
		    \Bigl(\frac{1}{\epsilon}+L_b\Bigr)\,
			(\Gamma_1)^{abcd}_{ABCD}	\;,\\[5pt]
	(Q6)&=-8\,\frac{\lambda^2}{16\pi^2}\,
		    \Bigl(\frac{1}{\epsilon}+L_f\Bigr)\,
	    \left[(\Gamma_1)^{abcd}_{ABCD}+(\Gamma_2)^{abcd}_{ABCD}\right]
	    ,\\[9pt]
	(Q7)&=-4\lambda\> \delta Z^{(1)}\, (\Gamma_1)^{abcd}_{ABCD}	\;.
\end{align}
\end{subequations}
Diagram $(Q6)$ requires a special attention because the $X_p$ and $Y_p$ scalars
have different Yukawa couplings.
Considering the various cases
(such as $\langle X^a_{p}\,X^b_{q}\,Y^c_{r}\,Y^d_{s}\rangle$)
and taking care of the appropriate permutations, one finds
the simple expression above.
Inserting the previously determined one-loop
scalar wavefunction renormalization factor $\delta Z^{(1)}$
[from Eq.~(\ref{Z})] into the result $(Q7)$ gives
\begin{equation}
	(Q7)=4\,\frac{\lambda^2}{16\pi^2\epsilon}\,
		(1+\alpha) \, (\Gamma_1)^{abcd}_{ABCD}	\;.
\end{equation}

Combining all terms, we find that the
amputated 1PI scalar 4-point function, at zero momentum, is
\begin{align}\label{quarticVSYM}
	\langle
	    \Phi_A^a
	    \Phi_B^b
	    \Phi_C^c
	    \Phi_D^d
	\rangle_{\rm 1PI}
	&=
	(\Gamma_1)^{abcd}_{ABCD} \,
	\Bigl(
	    - 2\, \lambda\,
	    + \frac{\lambda^2}{4\pi^2} [(1{-}\alpha) L_b - 2 L_f]
	\Bigr)
\nonumber\\
	&+
	(\Gamma_2)^{abcd}_{ABCD} \>
	\frac{\lambda^2}{2\pi^2} \>
	    [ L_b - L_f]
	+ \mathcal O(\lambda^3) \;.
\end{align}
Rescaling the field according to
relations (\ref {eq:rescale}) and (\ref {eq:Z3})
removes the remaining gauge fixing parameter dependence
and produces the equivalent ESYM$_3$ 1PI 4-point function,
\begin{align}\label{quarticV}
	\langle
	    \tilde\Phi_A^a
	    \tilde\Phi_B^b
	    \tilde\Phi_C^c
	    \tilde\Phi_D^d
	\rangle_{\rm 1PI}
	&=
	-2\lambda T^2\, (\Gamma_1)^{abcd}_{ABCD}
	-(\ln4) \, \frac{\lambda^2T^2}{\pi^2}
	\left[ (\Gamma_2)^{abcd}_{ABCD} - (\Gamma_1)^{abcd}_{ABCD} \right]
	+ \mathcal O(\lambda^3) \;.
\end{align}
The first term is the vertex which arises from
the commutator squared interaction directly inherited from the
full theory,
while the order $\lambda^2$ correction comes
out to be remarkably simple.
Both terms must be produced as tree level vertices
in the three-dimensional effective theory.
The required quartic interaction terms in the ESYM$_3$
effective Lagrangian are
\begin{equation}
	\coeff{1}{2} g_3^2\, \tr(i[\tilde\Phi_A,\tilde\Phi_B])^2
	+\frac{\ln 2}{2\pi^2}\, \frac{g_3^4}T\,
	\tr(\tilde\Phi^a_A V_a \tilde\Phi^b_B V_b)^2 \,,
\label{eq:quarticterms}
\end{equation}
with $g_3^2 \equiv g^2 T$.
All gauge fixing parameter dependence
and $1/\epsilon$ poles have canceled,
as required.
The $\ln (\Lambda^2/T^2)$ terms in $L_b$ and $L_f$ have also dropped out.
This was also required, since $\Nfour$ SYM at zero temperature
is a conformal theory in which the coupling does not run,
and turning on a temperature does not change the short distance behavior.

\section{Background Field Method}\label{BGField}

We wish to evaluate the complete
one-loop effective potential for the scalar fields
using background field techniques.
We will restrict our analysis to the special case where the background scalar
fields take values along flat directions of the tree-level potential.
This is sufficient for our purposes, and
simplifies the calculation.
We carry out the calculations in flat space, and neglect
chemical potentials.
Under our assumptions that $1/R^2$ and $\mu^2$ are both of order
$\lambda T^2$,
the tree-level contributions from the curvature and chemical potential
induced mass terms will be comparable to one-loop corrections to the
effective potential computed in flat space with $\mu = 0$.
Including curvature and chemical potential corrections in the
evaluation of the one-loop potential would change the result
by an amount comparable to neglected two-loop contributions.

In the flat space Lagrangian, shown in Eq.~(\ref{fullLagrangian}) but without
mass and chemical potential terms,
we shift the scalar fields by constants,
\begin{equation}\label{scalarshift}
	\Phi_A\rightarrow \bar\Phi_A + \sigma_A
								\;,
\end{equation}
with $\bar\Phi_A$ constant fields taking
values in the flat directions,
and $\sigma_A$, are arbitrary
fluctuations about the constant background.
The background fields (viewed as $\Nc \times \Nc$ matrices)
are simultaneously diagonal
(up to an irrelevant gauge transformation).

Among the quadratic terms in the shifted Lagrangian,
there are terms which mix the gauge and scalar fields.
These may be eliminated
by employing the $R_\xi$-gauge with gauge-fixing term
\begin{equation}\label{eq:R_xi}
    \frac{1}{\xi}\tr\!
    \left\{(\partial_\mu A_\mu-i\xi \, g[\sigma_A,\bar\Phi_A])^2\right\}
									\;.
\end{equation}
We further set $\xi=1$ to simplify the computation, as
this choice eliminates off-diagonal quadratic terms which
mix $\sigma_A$ and $\sigma_B$ for $A \ne B$.
We denote the matrix components of the fields by indices $m$ and $n$
so that,
for example,
$(\sigma_A)_{mn}\equiv \sum_a\sigma_A^a(T^a)_{mn}$.

Let $\lambda^m_A$ denote the $m$th eigenvalue of the constant background
field $\bar\Phi_A$ and define, for convenience,
\begin{equation}\label{DefLambdaDefM}
	M^2_{mn} \equiv g^2\sum_A \> (\lambda^{m}_A - \lambda^n_A)^2	
								\;.
\end{equation}
Note that $m$ and $n$ are
not to be summed over in the definition
of $M_{mn}^2$.
The resulting quadratic terms in the shifted Lagrangian are
\begin{align}
	\sum_{m,n}\bigg[
	&(A_\mu)^*_{mn}\left(-\partial^2+M_{mn}^2\right)(A_\mu)_{mn}
	+(\sigma_A)^*_{mn}
	\left(-\partial^2+M_{mn}^2\right)(\sigma_A)_{mn}
								\nonumber\\
	&+(\bar{\psi_i})_{mn}
	\left[\delta_{ij} \, i \Slash{\partial}
		- g\left\{\alpha^p_{ij} +i\gamma_5\, \beta^p_{ij} \right\}
		    (\lambda^{m}_p{-}\lambda^{n}_p)
		\right]
	(\psi_j)_{mn}
						\bigg]
								\,.
\end{align}
In carrying out the Gaussian integrals, one finds that
the logarithms of the functional determinants from
all the three contributions
simplify to the generic form
\begin{equation}
	I=\sumintp\ln(p^2+M_{mn}^2)
								\;,
\end{equation}
where the sum-integral is defined in Eq.~(\ref{defsumint}).
This integral is standard after Arnold and Espinosa
\cite{Arnold:1992rz}, and yields
\begin{subequations}
\begin{align}
	\frac{I_\text{bose}}{\pi^2T^4}
	    &=-\coeff 1{45}
	      +\coeff 13 \Big(\frac{M_{mn}^2}{4\pi^2T^2}\!\Big)
	      -\coeff 12 \Big(\frac{1}{\epsilon} {+} L_b \!\Big) \!
			 \Big(\frac{M_{mn}^2}{4\pi^2T^2}\!\Big)^2
	      - \sum_{l=3}^{\infty}8 \,
		    \frac{(2l{-}5)!!}{(2l)!!} \, \zeta (2l{-}3)
		    \Big(\frac{-M_{mn}^2}{4\pi^2T^2}\!\Big)^l \,,
\\
	\frac{I_\text{fermi}}{\pi^2T^4}
	    &=\coeff 7{360}
	      -\coeff 16 \Big(\frac{M_{mn}^2}{4\pi^2T^2}\!\Big)
	      -\coeff 12 \Big(\frac{1}{\epsilon} {+} L_f \!\Big) \!
			 \Big(\frac{M_{mn}^2}{4\pi^2T^2}\!\Big)^2
	      - \!\sum_{l=3}^{\infty}(4^l{-}8)
			\frac{(2l{-}5)!!}{(2l)!!} \,\zeta (2l{-}3)
			\Big(\frac{-M_{mn}^2}{4\pi^2T^2}\!\Big)^l ,
\end{align}
\end{subequations}
where $I_\text{bose}$ and $I_\text{fermi}$ are contributions from the bosons
and fermions, respectively.
In these results,
$\mathcal O(\epsilon)$ terms have been discarded and
the quantities $L_b$ and $L_f$ are defined in Eq.~(\ref{LbLf}).

For each color component,
the gauge field, six scalar fields, and gauge fixing ghosts contribute
$(4+6-2)=8$ times $\coeff \beta 2 I_{\rm bose}$
to the (three dimensional) effective potential, while
the fermions contribute $-8$ times $\coeff \beta 2 I_{\rm fermi}$.
The resulting effective potential,
for background fields lying along flat directions, is thus
\begin{align}\label{effp}
	\overline V(\bar\Phi_A)
	=
	\half \pi^2T^3\sum_{m,n}\bigg[\,
	-\frac 13
	&+\left(\frac{M_{mn}^2}{\pi^2T^2}\right)
	+(\ln 2)\left(\frac{M_{mn}^2}{\pi^2T^2}\right)^2
								\nonumber \\
	&+\sum_{l=3}^{\infty} \> 8 \, (1-4^{-l+2}) \,
	\frac{(2l{-}5)!!}{(2l)!!} \, \zeta (2l{-}3)
	\left(-\frac{M_{mn}^2}{\pi^2T^2}\right)^{l} \, \bigg]
								\;.
\end{align}
The $1/\epsilon$ poles and all dependence on the
arbitrary renormalization scale has canceled, as required
for a physical quantity.

We would like to express the result (\ref {effp})
in terms of the original fields $\bar\Phi_A$.
To do so, note that the adjoint representation
of the field,
$
	(\bar\Phi_A^{\text{adj}})_{ab}\equiv i\bar\Phi_A^{b}f^{abc}
$,
may also be written as
\begin{equation}\label{DefAdj}
	(\bar\Phi_A^{\text{adj}})_{ab} = 2\tr (T_a[\bar\Phi_A,T_b])
								\;.
\end{equation}
Since the fields $\bar\Phi_A$ are assumed to take values in flat directions,
a Cartan subalgebra may chosen for which $\bar\Phi_A$ is
diagonal with the entries $\lambda_A^m$.
Therefore
\begin{align}
	\tr (\bar\Phi_A^{\text{adj}}\bar\Phi_B^{\text{adj}})
	&=2\sum_a\tr (T_a[\bar\Phi_A,[\bar\Phi_B,T_a]])
\nonumber\\
	&=2\sum_{a,m,n}(T_a)_{nm}
		(\lambda_A^{m}{-}\lambda_A^{n})
		(\lambda_B^{m}{-}\lambda_B^{n})(T_a)_{mn}
\nonumber\\
	&= \sum_{m,n} \>
		(\lambda_A^{m}{-}\lambda_A^{n})
		(\lambda_B^{m}{-}\lambda_B^{n}) \,,
\end{align}
where the last step used the identity
$
	\sum_a(T_a)_{ij}(T_a)_{kl}=\frac{1}{2}\,(\delta_{il}\delta_{jk}
			-\frac{1}{N_c}\,\delta_{ij}\delta_{kl})
$.
Hence
$
	\tr [g^2 \sum_A \bar\Phi_A^{\text{adj}}\bar\Phi_A^{\text{adj}}]
	= g^2 \sum_{A,m,n}(\lambda_A^{m}{-}\lambda_A^{n})^2
	= \sum_{m,n} M_{mn}^2
$.
Since the sum of the eigenvalues $\lambda_A^m$ must vanish
(because $\bar\Phi_A$ is traceless), this may also be written in the form
\begin{equation}\label{quadtranslation}
	\half g^2T^2\tr (\bar\Phi_A^{\text{adj}}\bar\Phi_A^{\text{adj}})
	= \half T^2 \sum_{m,n}M_{mn}^2
	= g^2 T^2 \Nc \sum_{A,m} (\lambda_A^m)^2
	=\lambda T^2\tr(\bar\Phi_A)^2			\;,
\end{equation}
which shows that the quadratic part of the effective potential (\ref{effp})
agrees with the previously derived thermal mass correction
(\ref{MassCorrection}).
[Note that, to the lowest order,
the rescaling from $\Phi_A$ to $\tilde\Phi_A$ in
(\ref{eq:rescale}) is just a factor of $\sqrt T$.]
More generally, we have
\begin{equation}\label{translation}
    \tr \Big[
	\Big(g^2\sum_A\bar\Phi_A^{\text{adj}}\bar\Phi_A^{\text{adj}}\Big)^l
	\Big]
    =
    \sum_{m,n}\> \left(M_{mn}^2\right)^l \;,
\end{equation}
so the effective potential (\ref {effp})
(along flat directions) may be expressed in terms of
adjoint representation traces as shown in Eq.~(\ref {eq:Vflat}).
In particular, the quartic term of (\ref{effp}) is
$
	\frac \beta {2 \pi^2} \,  (\ln 2) \, g^4
	\tr [(\bar\Phi^{\text{adj}}_A\bar\Phi^{\text{adj}}_A)^2]
$
and this agrees, for commuting fields, with the diagrammatic result
(\ref{eq:quarticterms}).%
\footnote{ 
The meticulous reader 
may be puzzled why this result agrees with Eq.~(\ref{ESYMquartic}),
without taking into account the order $\lambda$ term in the
wavefunction renormalization $Z_3$ [given in Eq.~(\ref{eq:Z3})].
As seen explicitly in Appendix \ref{QuarticSubsection},
this rescaling leads to a term proportional to the
tree-level interaction and hence
vanishes for fields lying along flat directions.
}




\end{document}